\documentclass{aa}
\usepackage{graphicx}
\newcommand{\SSS}{\scriptscriptstyle}
\newcommand{\Rl}{R_{\SSS L1}}
\newcommand{\Md}{\dot{\cal M}}
\newcommand{\Msol}{{\cal M}_\odot}
\newcommand{\SC}{\scriptsize}
\newcommand{\Teff}{T_{\mbox{\SC eff}}}
\newcommand{\TeffA}{T_{{\mbox{\SC eff}},{\rm A}}}

\newcommand{\gsim}{{\textstyle{\; \lower 0.7ex\hbox{$>$}\; 
\atop \raise-0.1ex\hbox{$\sim$}}}}
\newcommand{\lsim}{{\textstyle{\; \lower 0.7ex\hbox{$<$}\; 
\atop \raise-0.1ex\hbox{$\sim$}}}}

\newcommand{\gs}{gs$^{-1}$}
\begin{document}
   \title{UU Aqr from high to low state}

   \author{Sonja~Vrielmann\inst{1}
          \and
	   Raymundo Baptista\inst{2}}
   \offprints{Sonja~Vrielmann}

   \institute{Department of Astronomy, University of Cape Town, Private Bag,
		Rondebosch, 7700, South Africa\\
		\email{sonja@pinguin.ast.uct.ac.za}
		\and
		Departamento de Fisica, Universidade Federal de Santa Catarina, Campus Trindade, 88040-900 Florianspolis - SC, Brazil\\
		\email{bap@astro.ufsc.br}
             }

   \date{Received; accepted}

\abstract{
In this paper we present Physical Parameter Eclipse Mapping (PPEM) of
UBVRI eclipse light curves of UU~Aqr from high to low states. We used
a simple, pure hydrogen LTE model to derive the temperature and
surface density distribution in the accretion disc. The
reconstructed effective temperatures in the disc range between
9\,000~K and 15\,000~K in the inner part of the disc and below
7\,000~K in the outer parts. In the higher states it shows a more or
less prominent bright spot with $\Teff$ between about 7\,000~K and
8\,000~K.\\
The inner part of the disc ($R < 0.3\Rl$) is optically thick at all
times, while the outer parts of the disc up to the disc edge
($0.51\pm0.04\Rl$ in the high state and $0.40\pm0.03\Rl$ in the low
state) deviate from a simple black body spectrum indicating that either
the outer disc is optically thin or it shows a temperature inversion
in the vertical direction.\\
While during high state the disc is variable, it appears rather stable
in low state. The {\em variation during high state} affects the size
of the optically thick part of the disc, the white dwarf or boundary
layer temperature and the uneclipsed component (originating in a disc
chromosphere and/or cool disc wind), while the actual size of the disc
remains constant.  The {\em difference between high and low state} is
expressed as a change in disc size that also affects the size of the
optically thick part of the disc and the presence of the bright
spot.\\
Using the PPEM method we retrieve a distance for UU~Aqr of 207$\pm$10~pc,
compatible with previous estimates.
}

\maketitle
%
%________________________________________________________________

\section{Introduction}
Cataclysmic variables (CVs) are close interacting binaries in which
the white dwarf, the primary component, accretes matter from a red
dwarf companion (for extensive reviews on CVs see Warner 1995 or
Hellier 2001). Depending on the strength of the magnetic field of the
white dwarf and the mass accretion rate from the secondary the system
geometry and physics of accretion varies. We deal in this paper with a
white dwarf primary that has a negligible magnetic field. In this case
the matter drawn from the secondary is accreted through an accretion
disc. A relatively high accretion rate as in the nova-like UU~Aqr
leads presumably to a steady state disc that does not show outbursts
like low accretion rate dwarf novae.

UU~Aqr is a relatively bright CV with a magnitude of 13.5 and an
orbital period of $3^h56^m$ that was first reckognized as an eclipsing
CV by Volkov, Shugarov \& Seregina~(1986). This makes it a relatively
easy target to observe and indeed it has been observed in the IR (Dhillon et
al. 2000, Huber \& Howell 1999), in the UV (White et al.\ 1995) as
well in the optical wavelength range. Optical photometry was reported
by Honeycutt, Robertson \& Turner (1998) and Baptista, Steiner \&
Cieslinski (1994, hereafter BSC) while optical spectroscopic data were
analysed by Baptista et al.\ (2000), Kaitchuck et al.\ (1998), Hoard
et al.\ (1998), Diaz \& Steiner (1991), and Haefner (1989).  It has
not been detected in the ROSAT PSPC All Sky Survey (Verbunt et al.\
1997) like two thirds of the nova-like systems in Verbunt et al.'s
list. These observations have led to relatively well-defines system
parameters, like the inclination angle $i=78^\circ$, mass ratio $q =
0.30$ and white dwarf mass ${\cal M}_{wd} = 0.67\Msol$.

Since it shows single peaked emission line profiles, Baptista, Steiner
\& Horne (1996, hereafter BSH) first proposed that UU~Aqr is a member
of the SW~Sex type CVs. Other features of these objects are
phase-dependent absorption components, hardly eclipsed emission lines,
and large phase shifts between photometric and spectroscopic
ephemerides.  Doppler maps are bright in the lower left quadrant and
the eclipses are V shaped, indicating a flat temperature profile
(e.g. Warner 1995). Furthermore, these systems do not usually show the
rotational disturbance of an eclipsed accretion disc.

UU~Aqr shows all these features (for a summary see Baptista et al.\
2000). Haefner (1989) already found the emission line profiles to
be variable and states that it ``indicates several sources of emission
within the system: symmetrical profiles as well as strong asymmetries
or even double peaks are present''. The variability could be caused by
a variable disc wind which sometimes is substantially reduced allowing
the double peaked emission from the disc to be seen.
Kaichuck et al.\ (1998) and Baptista
et al.\ (2000) suggest that the emission lines originate in a
chromosphere and a disc wind and/or that the disc has a disc anchored
magnetic propeller in which the gas is ejected in all
directions. Hoard et al.\ (1998) propose a detailed model for the
emission and absorption sites in the disc.

BSH discovered that UU~Aqr changes between high and low states on a
timescale of a couple of years and that the main difference is the
presence of a bright spot in the high state and lack of it in the low
state. In addition, Honeycutt et al.\ (1998) observed so-called
``stunted'' outbursts which have the duration of a dwarf nova
outburst, but only small amplitudes of less than a magnitude (compared
to 2-3 mag in dwarf novae outbursts).

The aim of this paper is to investigate quantitatively the difference
between high and low states and the variation within those states. We
analysed BSH's eclipse light curves individually and combined them in
six sub-states to minimize the influence of flickering and flaring.

A second objective was to determine the distance to UU~Aqr using the
PPEM method and compare it to previous estimates. BSC determined a
distance of 270$\pm$ 50~pc by fitting the white dwarf flux. In BSH
they used a method similar to cluster main sequence fitting and
arrived at a distance of 200$\pm$30~pc.

\section{The data and system parameters}
The data set was previously presented by BSC and BSH. The 37 eclipses
of UU~Aqr observations were taken with the 0.6m and the 1.6m
telescopes of the Laboratorio Nacional da Astrofisica (LNA/CNPq) in
Brasil. Further details about the data aquisition and reduction can be
found in BSC.

Apart from analysing only the high and low state data, we also
analysed the individual light curves with the PPEM method. The aim was
to retrieve more detail using the PPEM algorithm, since this method
allows us to obtain the distributions of temperature and surface
density over the disc using the information of the five colours
simultaneously.

\section{The PPEM analysis}
The Physical Parameter Eclipse Mapping method was presented by
Vrielmann, Horne \& Hessman (1999, hereafter VHH). An overview of
applications of this method to various systems is given in Vrielmann
(2001). Vrielmann, Stiening \& Offutt (2002a) and Vrielmann, Hessman
\& Horne (2002b) show applications to V2051~Oph and HT~Cas,
respectively.  We will give here only a very short description of the
method, the interested reader is referred to the above articles
for more details.

The idea is based on the Eclipse Mapping developed by Horne (1985)
which uses the eclipse profile in order to reconstruct the light
distribution in the eclipsed source, i.e.\ the white dwarf and the
accretion disc. Hereby, a Maximum-Entropy-Method (MEM) helps to choose
the distribution (or map) with the least structure still compatible
with the observed data.

 PPEM goes a step further in that it uses multi-band light curves in order
to reconstruct distributions of physical parameters, e.g.\ the
temperature and surface density of the eclipsed object. This means
a spectral model for the emissivity of these objects has to be adopted.

In this study, we use white dwarf spectra and a simple, pure hydrogen
spectrum for the accretion disc as described in the next Section.
Future studies (see Vrielmann, Still \& Horne 2002c) will include more
sophisticated disc model, like those computed by Hubeny
(1991).

\subsection{The spectral model used for PPEM}
\label{spec_model}

For the reconstruction of the physical parameters we used a uniform,
pure hydrogen LTE slab including only free-free and bound-free
emission (as described in more detail in VHH):
\begin{equation}
\label{othin}
        I_\nu(T,\Sigma) = B_\nu(T) \cdot \left[1 -
        e^{-\frac{\tau_\nu}{\cos i}}\right]
\end{equation}
where $B_\nu(T)$ is the blackbody spectrum, $i$ the inclination angle
and $\tau_\nu$ the optical depth, calculated as:
\begin{equation}
        \tau_\nu = \kappa_\nu \Sigma
\end{equation}
where $\kappa_\nu(\rho,T)$ is the mass absorption coefficient for
hydrogen, including atomic and H$^-$ bound-free and free-free
contributions.  The mass-density $\rho$ is calculated from $\Sigma$
and $T$ using the vertical half-thickness ($c_s$: sound speed,
$V_{\mbox{\tiny Kepler}}$: keplerian velocity):
\begin{equation}
\label{eq_h}
        H(R,T) = \frac{c_s(T) R}{V_{\mbox{\tiny Kepler}}}
\end{equation}
For $\frac{\tau}{\cos i} \gg 1$, Eqn.~\ref{othin} transforms into the
optically thick case $I_\nu \approx B_\nu(T)$.  For the optically thin
case $\frac{\tau}{\cos i} \ll 1$, Eqn.~\ref{othin} reduces to
$I_\nu(T) \approx \frac{\tau}{\cos i} \cdot B_\nu(T)$.

We choose the parameters temperature $T$ and surface density $\Sigma$
to be reconstructed. From the reconstructed maps in $T$ and $\Sigma$
we can e.g.\ calculate the effective temperatures, the optical depth,
or the Balmer Jump strength, i.e.\ the ratio
$I_\nu(T,\Sigma)/B_\nu(T)$ (see Section~\ref{maps}).

Although this model is simple, it is very useful, because it allows us
to diffentiate between optically thick and optically thin parts, or
more generally, disc regions with emissivity resembling or deviating
from the black body spectrum. Furthermore, for typical temperatures
and surface densities in real discs, LTE and the pure hydrogen
assumption are good approximations. 
On the other hand, the opacity in the cooler regions of real disc will 
be typically much higher, leading to overestimated temperatures for 
disc regions cooler than about 6300~K.
For a more detailed discussion on the applicability of the
model see Vrielmann et al.\ (2002b).

Thus, deviations from BB emission (Balmer jump in emission or
absorption) are attributed to optical depth effects. If such deviations
are instead due to temperature gradients in the disc atmosphere, our
simple model will produce systematic errors. In realistic discs, it is
possible that the inner part of the disc shows a stellar-type
decreasing temperature with disc height while the outer regions show a
chromospheric-type temperature inversion. If this is the case our
simple model will fail.

\subsection{PPEM distance estimates}

%%%%%%%%%%%%%%%%%%%%%%%%%%%%%%%%%% FIGURE %%%%%%%%%%%%%%%%%%%%%%%%%%%%%%%%%%

   \begin{figure}
   \centering
   \includegraphics[width=8cm]{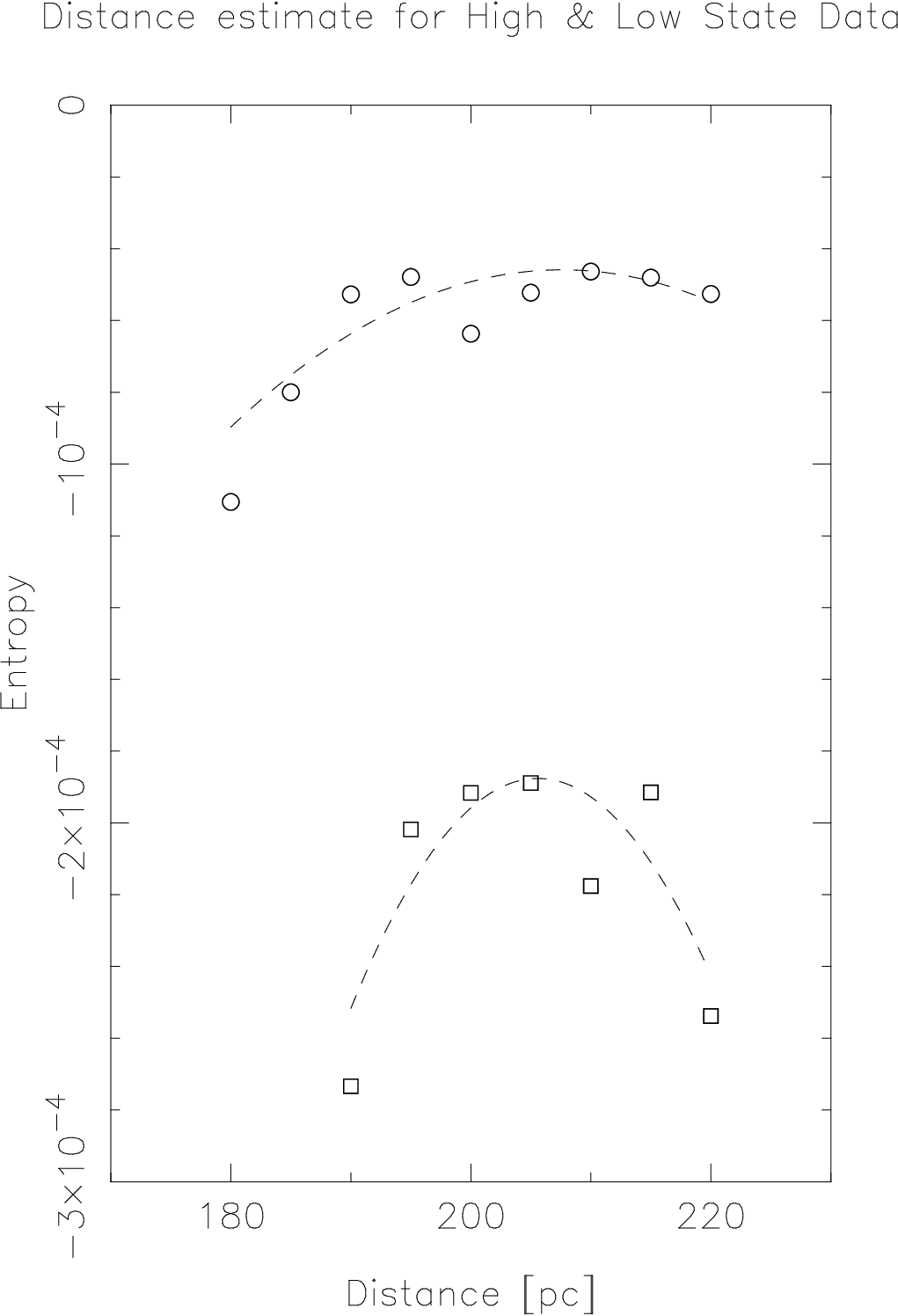}
      \caption{The entropy of the reconstructions (a measure of
smoothness) as a function of the assumed distance. For each trial
distance the data were fitted with $\chi^2/N = 2$ (high state, squares) and
1.25 (low state, circles). The dashed lines are parabolic fits to the data, they
peak at 206~pc and 208~pc, respectively.}
\label{fig_dist}
   \end{figure}

%%%%%%%%%%%%%%%%%%%%%%%%%%%%%%%%%% FIGURE %%%%%%%%%%%%%%%%%%%%%%%%%%%%%%%%%%

Using the PPEM method we determined a new distance estimate for
UU~Aqr. The distance-entropy relation for the fits to the high and low
state data is plotted in Fig.~\ref{fig_dist} and shows clear peaks at
206~pc and 208~pc. Due to lack of any true constraint to weigh the two
values we take the simple average of the distance (207~pc). In
Vrielmann et al. (2002b) we show a test case using an
artifical non-axisymmetric disc and derive a distance 2.5~\% larger
than the assumed true distance. We have not included reddening in these
calculation, however, as BSH point out, the reddening is nearly
negligible and would change the distance by at most 5~pc. We estimate
therefore the total error of the distance to 10~pc with the assumption
that our spectral model describes the true emissivity reasonable well.
For the following calculations we used the trial distance (and BSH's
distance) of 200~pc which should give basically identical results.

Our distance estimate is in very good agreement with BSH's value of
$200\pm30$~pc using a method similar to cluster main sequence fitting
for the accretion disc, however, it is only marginally compatible with
their distance derived from a fit to the white dwarf flux by BSC. As
BSC remark, they may have over-estimated the radius of the white
dwarf and included (part of) the boundary layer in its estimate.

This again shows that the PPEM method is in principle a very good
method to estimate distances of real objects. However, as described in
Vrielmann et al.\ (2002b) in the case that the disc has
patches that lead to a covering factor less than unity the PPEM
distance would be overestimated. The fact that our estimate agrees
with BSH's and is smaller (rather than larger) than the white dwarf
estimate by BSC leads to the conclusion that the disc is probably not patchy.

\subsection{The light curves}

%%%%%%%%%%%%%%%%%%%%%%%%%%%%%%%%%% FIGURE %%%%%%%%%%%%%%%%%%%%%%%%%%%%%%%%%%
\begin{figure*}
   \centering
   \includegraphics[width=57mm]{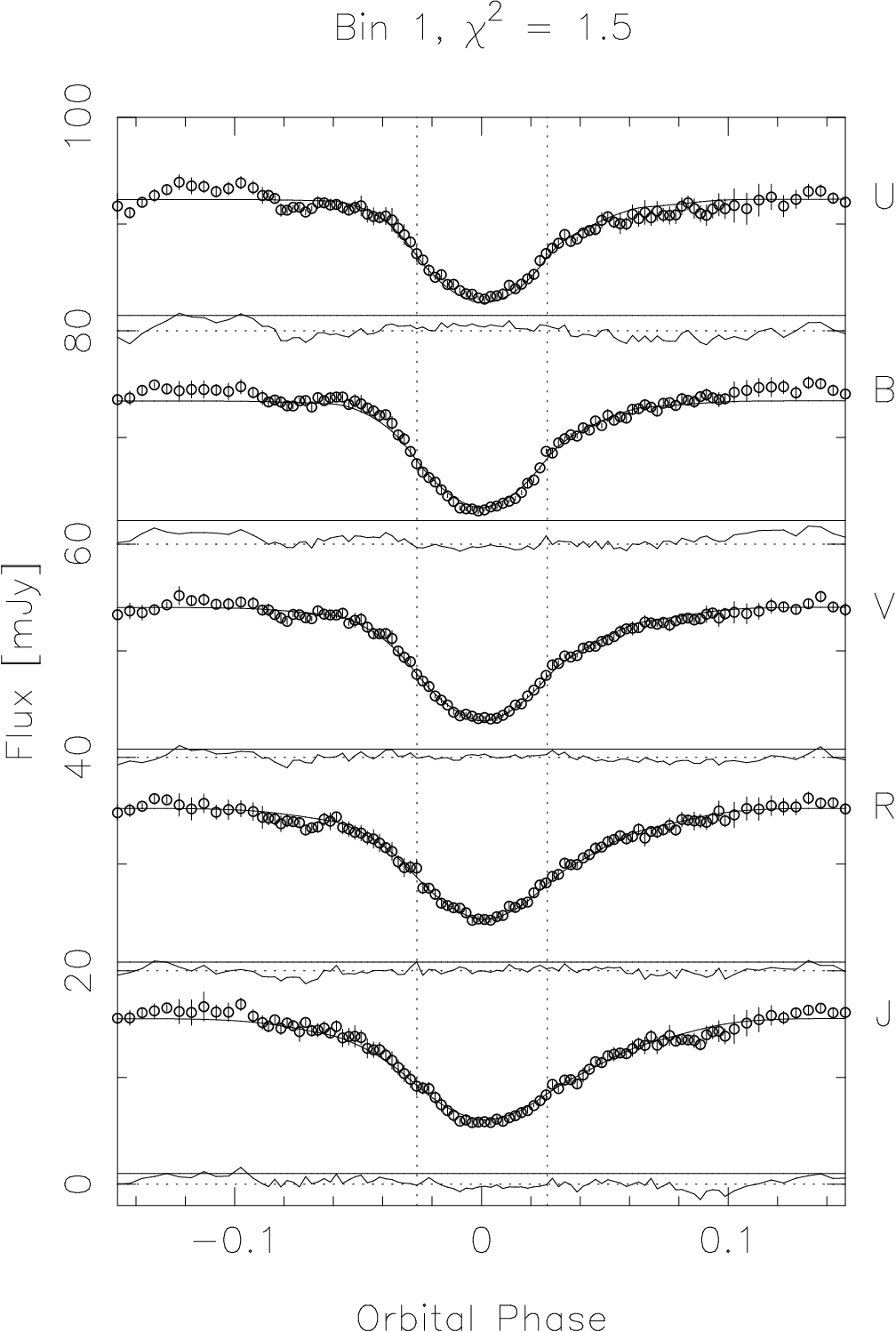}
   \includegraphics[width=57mm]{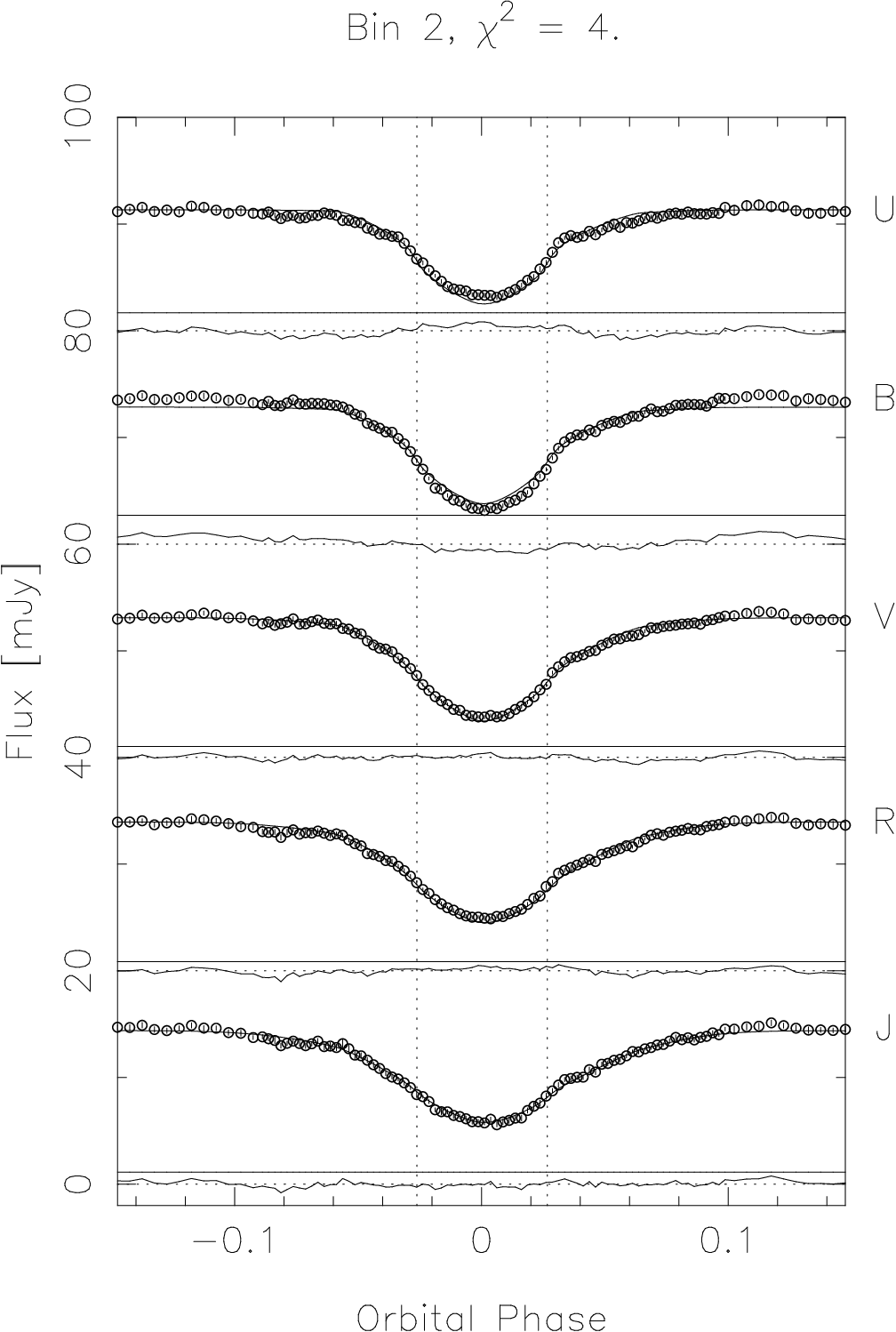}
   \includegraphics[width=57mm]{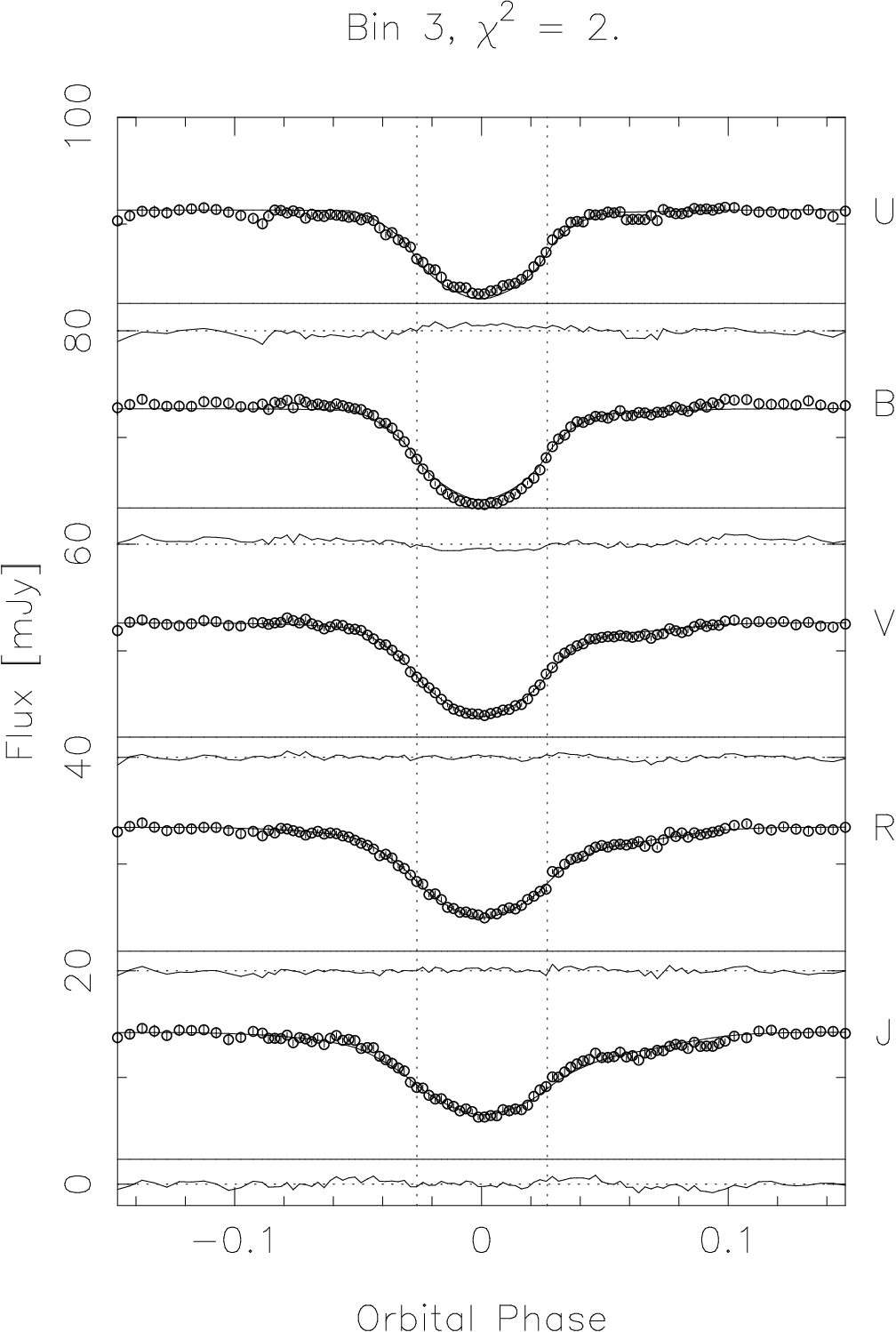}

\vspace*{3ex}
   \includegraphics[width=57mm]{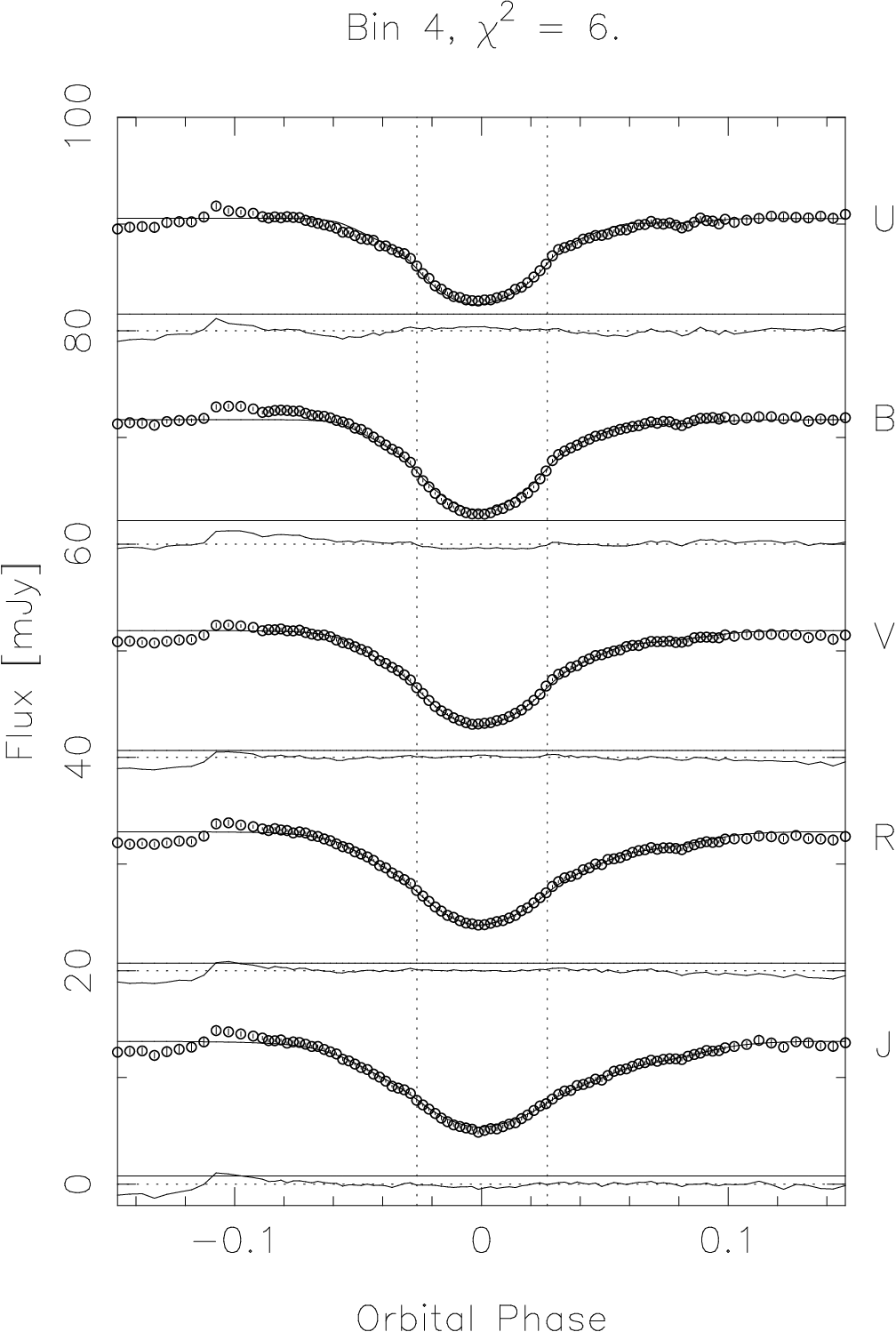}
   \includegraphics[width=57mm]{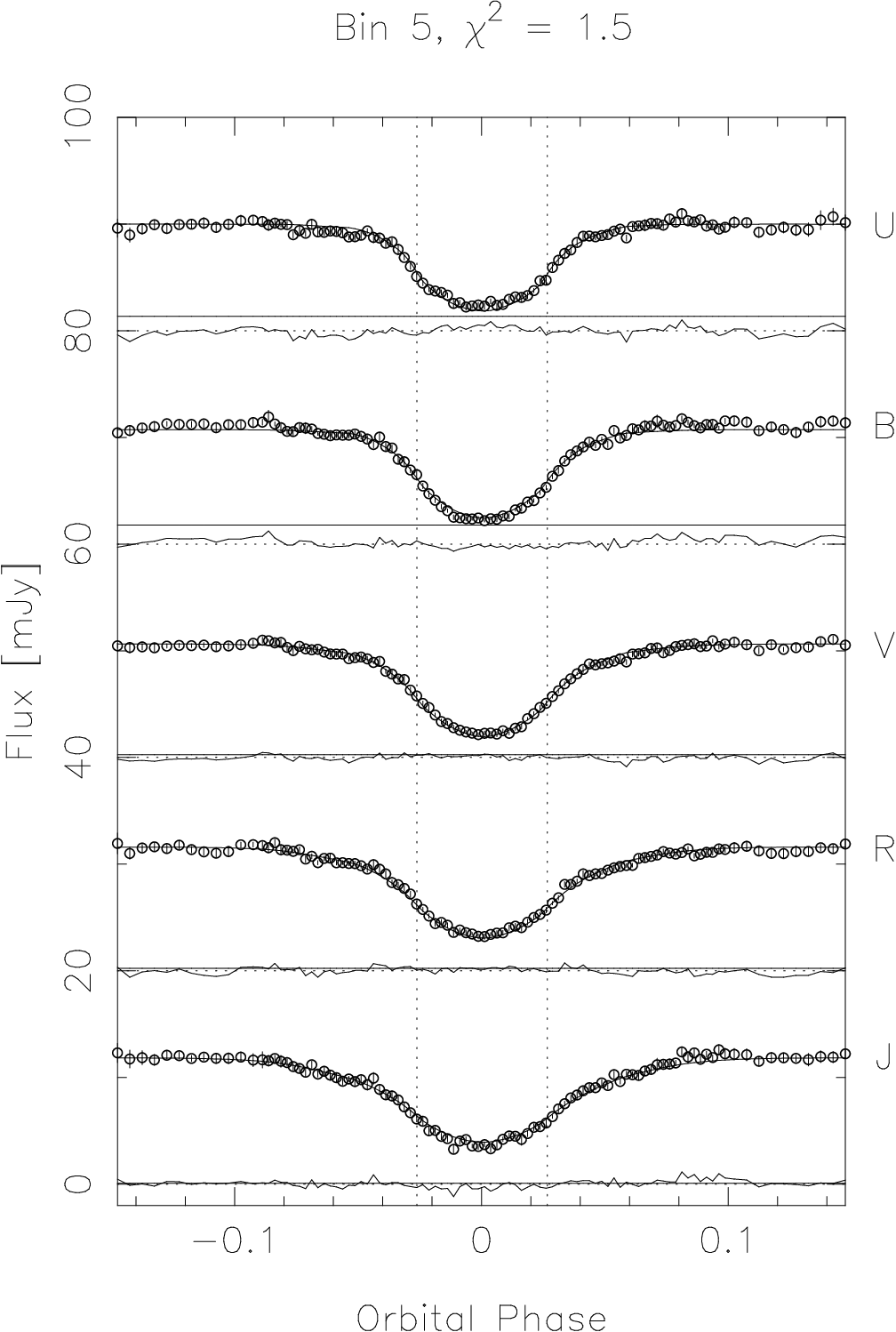}
   \includegraphics[width=57mm]{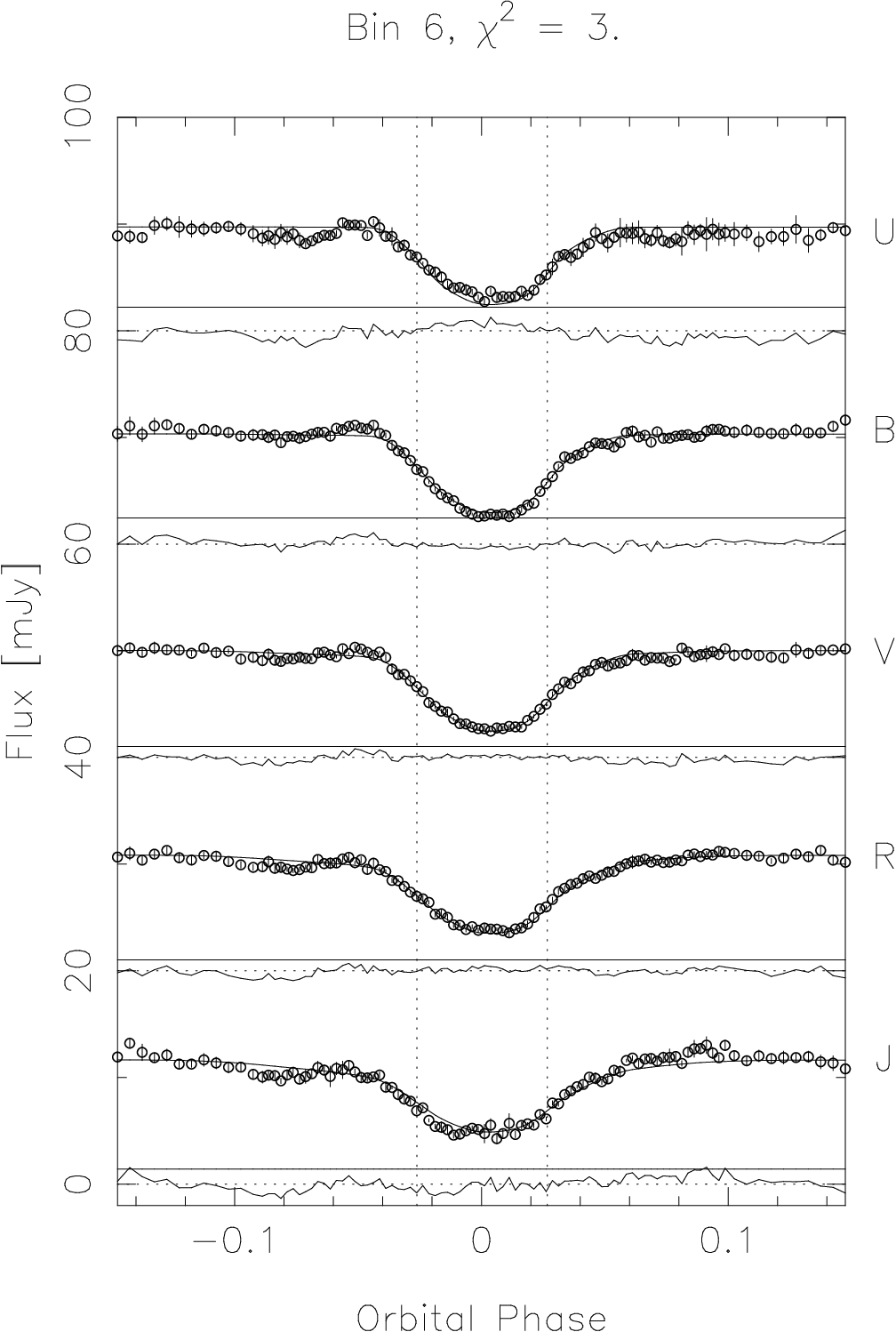}
\caption{The six binned light curves in UBVRI with the fits,
the residuals and the uneclipsed components.}
\label{fig_lcvs}
\end{figure*}
%%%%%%%%%%%%%%%%%%%%%%%%%%%%%%%%%% FIGURE %%%%%%%%%%%%%%%%%%%%%%%%%%%%%%%%%%

While BHS have binned the light curves by year of observation and then
separated into high and low states, we rearranged the light curves
according to decreasing out-of-eclipse flux in V. Since this did not
yield a clear separation into two different states, i.e.\ the high and
low state, instead, we recognized six categories and binned them accordingly
into 6 sub-states (see Tab~\ref{tab_sep}).

%%%%%%%%%%%%%%%%%%%%%%%%%%%%%%%%%% TABLE 1 %%%%%%%%%%%%%%%%%%%%%%%%%%%%%%%%%%%%
\begin{table}[ht]
   \centering
\caption{Eclipse numbers (see BSC) sorted into six bins (sub states) according to
the out-of-eclipse flux in V. Column 3 gives an estimate of the average flux in mJy
(errors are 0.3~mJy) and column 4 the corresponding magnitude.
\label{tab_sep}}
\vspace{1ex}
\begin{tabular}{clcc}
Bin & Eclipse No. & $<F>$ & $<m>$ \\ \hline
1 & 8592, 9025, 8512 & 13.9 & 13.55\\
2 & 8574, 8458, 8591, 8586, 8531 & 13.2 & 13.6\\
  & 10823 & & \\
3 & 8525, 8585, 8042, 8524, 8140 & 12.6 & 13.65\\
4 & 8030, 8530, 8518, 10822, 10828 & 11.8 & 13.7\\
  & 8024, 8744, 15200, 9031, 10829 & & \\ \hline
5 & 6739, 15212, 6745, 6744 & 10.8 & 13.8\\
6 & 15213, 15207 & 10.2 & 13.9\\
\hline
\end{tabular}
\end{table}
%%%%%%%%%%%%%%%%%%%%%%%%%%%%%%%%%%%%%%%%%%%%%%%%%%%%%%%%%%%%%%%%%%%%%%%%%%%%%%%

However, checking which light curves contributed to which sub-state, we
can rediscover the high and low states recognized by BHS: the high
state consisting of the sub-states 1 to 4 and the low state of
sub-states 5 and 6. The eclipses in sub-state 5 and 6 were taken in
different years or seasons, i.e. in 1988 and 1992, than the eclipses
in the other sub states (taken in 1989 and 1990, with the exception of
eclipse \#15200). A comparison with Honeycutt et al.'s (1998) data
shows that in July 1992 UU~Aqr was in a low state (however, not in the
lowest possible) with a V magnitude of about 13.8, while at other
times it varies around a V magnitude of 13.5. Some ten days later
(perhaps 30, definitely 50) UU~Aqr experienced a so-called ``stunted''
outburst\footnote{These ``stunted'' outburst have a small amplitude,
but usual length and were observed in nova-likes and old novae by
Honeycutt et al.\ (1998)}. Unfortunately, there is no more overlap
between the two data sets.  Apparently, the systems either rests in a
low state or varies within the brighter 4 sub-states.

We cannot exclude that the odd eclipse (e.g.\ \#8512 (sub-state 2?),
\#8531 (sub-state 3?), \#10823 (sub-state 3?)) might be missclassified
(to a sub-state too high), because the light curves usually only cover
the immediate eclipse and not enough signal before and after the
ingress and egress, respectively, to decide whether the flux just
preceding or following the eclipse was contaminated by a flicker or a
broad hump.

During the high state the system often changes in flux. The average
time scales for dropping from one sub-state to another is about five
orbital cycles or more. However, on two occasions the system dropped from
sub-state 1 to 4 within 6 cycles.  The rise from one sub-state to
another, even jumping an intermediate sub-state, however, can occur
much quicker, sometimes within one cycle.

In the case of the eclipses \#8530 and \#8531, during which the system
jumped two (or at least one) sub-states upwards, an inspection of the
original light curves clearly shows an increase in the redder
passbands (i.e. in BVRI). Furthermore, the eclipse \#8531 is much
shallower (also especially in the redder filters) and shows
indications of a bright spot. We must therefore conclude that the
system indeed changed significantly within this particular orbit,
possibly due to an increase in the uneclipsed component (disc wind).

%A PPEM analysis of these individual light curves show that the
%disc has no bright spot during eclipse \#8530 but a clearly visible one
%during eclipse \#8531. 

The binned light curves in UBVRI are shown in Fig.~\ref{fig_lcvs},
together with the fits, the residuals and the uneclipsed component.
All light curves show a shallower egress than ingress, indicative of a
bright spot, however, the asymmetry is strongest in the light curves
of the higher sub-states.

The amount of flickering and noise in the binned light curves is
inversely proportional to the number of individual light
curves in the respective bin number.

\section{Results}
\subsection{The reconstructed accretion disc}
\label{maps}

We reconstructed two-dimensional temperature $T$ and surface density
$\Sigma$ distributions of the accretion disc. As explained in
VHH, the
parameters can only be reconstructed with a certain accuracy,
depending on the value of each parameter (i.e.\ the location in the
parameter space). Usually, the temperature can be reconstructed with a
better accuracy than the surface density. (We strongly recommend the
reader to study the article VHH). As explained in Vrielmann et al.\
(2002a) and (2002b) the effective temperature, which also depends on
the total emission (the quantity that is fitted), is usually
relatively acurately reproduced. Furthermore, this parameter is also
usually dealt with in theoretical studies.

Another useful tool to investigate the accretion disc physics is the
parameter $I_{T,\Sigma}/I_{BB}$, where $I_{T,\Sigma}$ is the intensity
derived from the $T,\Sigma$ maps and $I_{BB}$ the black body
temperature derived from the $T$ map alone. This parameter measures
the deviations of the observed spectrum from a blackbody spectrum of
same temperature and, differently from $T$ and $\Sigma$, is a
model-independent quantity. It may also be understood 
as a measurement of the Balmer Jump strength. According to the adopted
spectral model, where this value reaches unity the disc is optically
thick and in regions where the emissivity deviates from the black body
it assumes values $0<I_{T,\Sigma}/I_{BB}<1$.  In extremely dim regions
(which can de facto mean outside the true disc radius) this ratio can
also reach nearly unity, because of the small values for
$I_{T,\Sigma}$ and $I_{BB}$ involved.

%%%%%%%%%%%%%%%%%%%%%%%%%%%%%%%%%% FIGURE %%%%%%%%%%%%%%%%%%%%%%%%%%%%%%%%%%
\begin{figure*}
   \centering
\includegraphics[width=170mm]{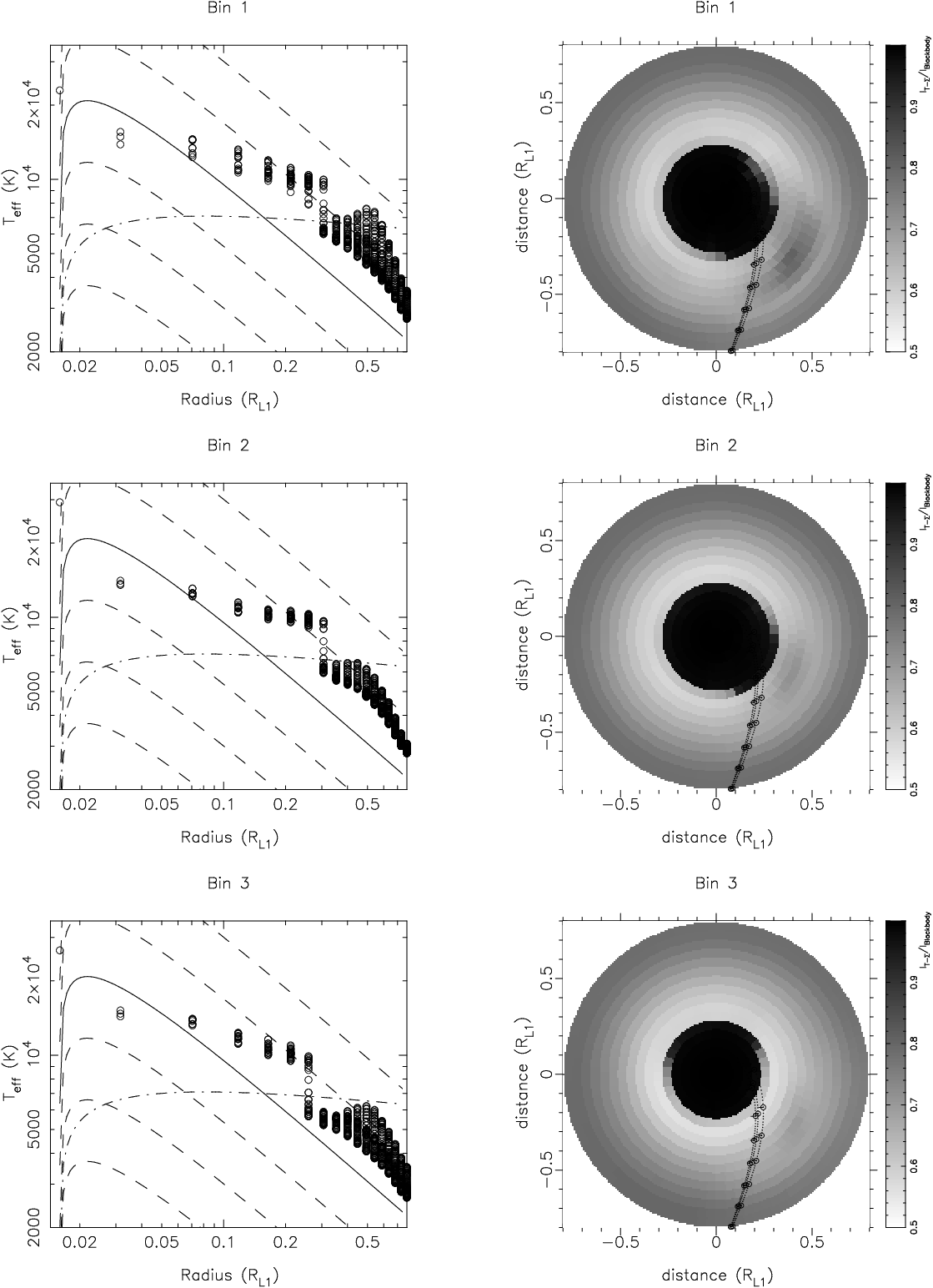}
\caption{{\em Left:} The effective temperature distribution derived
from the reconstucted temperature and surface density distributions
for the first three bins. The dashed lines give theoretical steady
state effective temperature distributions for mass accretion rates
$\log \Md = 13$ to 18, the one for $\Md = 10^{16}$\gs is drawn solid
for reference. The dashed-dotted line indicates the critical effective
temperature \protect{$\TeffA$} according to Ludwig, Meyer-Hofmeister
\& Ritter's (1994) critical mass accretion rate $\Md_{\rm A}$.  {\em
Right:} The Balmer Jump strength ($I_{T,\Sigma}/I_{BB}$) distribution
derived from the reconstucted temperature and surface density
distributions for the first three bins. A ratio of unity (black areas)
means the emitting material is optically thick.}
\label{fig_ti1}
\end{figure*}
%%%%%%%%%%%%%%%%%%%%%%%%%%%%%%%%%% FIGURE %%%%%%%%%%%%%%%%%%%%%%%%%%%%%%%%%%

%%%%%%%%%%%%%%%%%%%%%%%%%%%%%%%%%% FIGURE %%%%%%%%%%%%%%%%%%%%%%%%%%%%%%%%%%
\begin{figure*}
   \centering
\includegraphics[width=170mm]{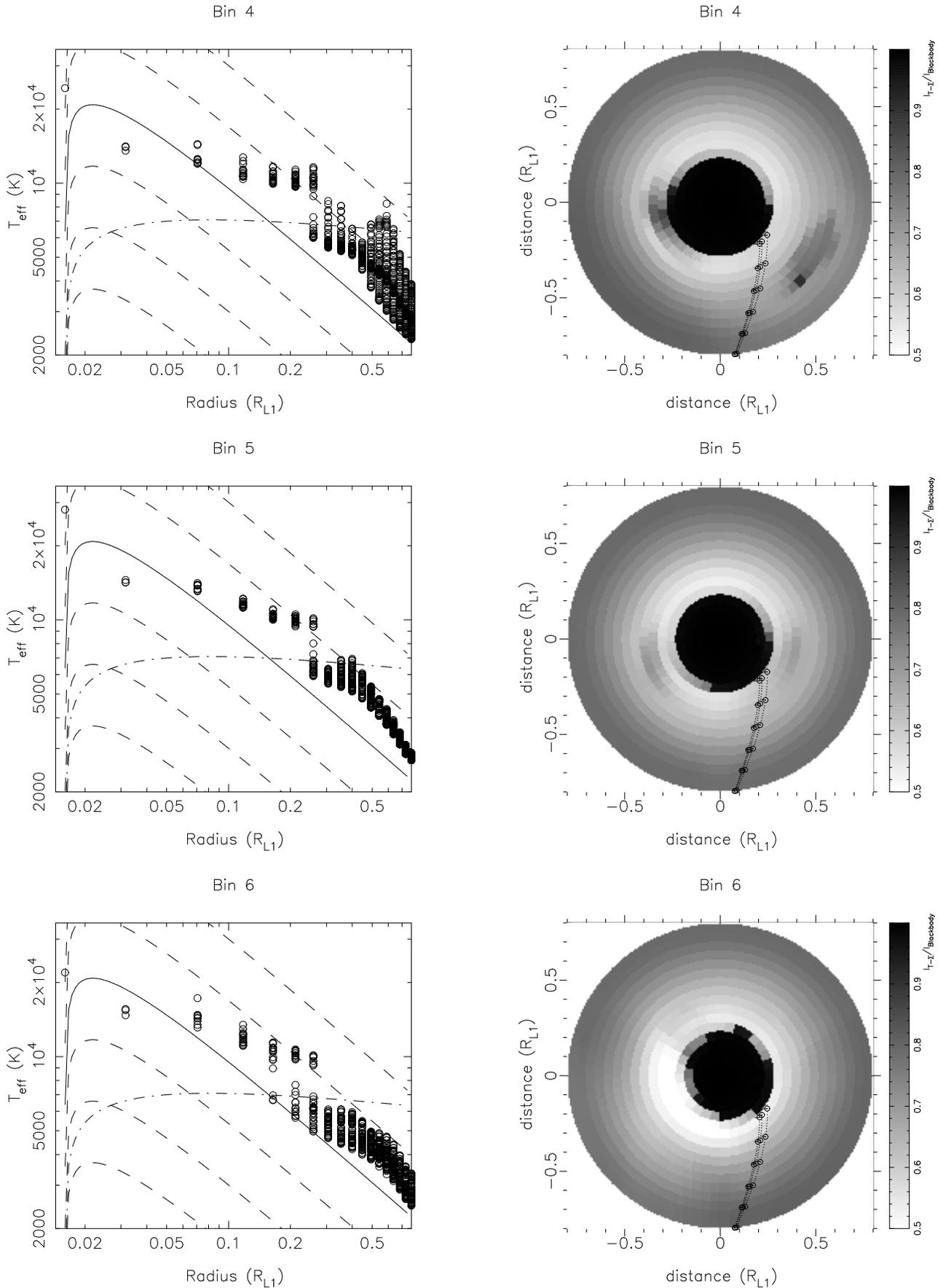}
\caption{As Fig.~\ref{fig_ti1} but for the last three bins.}
\label{fig_ti2}
\end{figure*}
%%%%%%%%%%%%%%%%%%%%%%%%%%%%%%%%%% FIGURE %%%%%%%%%%%%%%%%%%%%%%%%%%%%%%%%%%

Fig.~\ref{fig_ti1} and \ref{fig_ti2} show the effective temperature
and Balmer Jump strength ($I_{T,\Sigma}/I_{BB}$) distribution derived
from the reconstructed temperature and surface density
distributions. The pattern for $I_{T,\Sigma}/I_{BB}$ and the effective
temperature (not shown) are similar, indicating that hot regions are
emitting optically thick and cooler regions deviate from a black body
source of same temperature.
Most prominent is the clear transition between the optically thick
inner part of the disc with a radius of between 0.3$\Rl$ and 0.25$\Rl$
in the highest and lowest sub-states, respectively, and the outer,
optically thin regions (see also Section~\ref{sec_radii}).

We remark that the observed deviations from blackbody emission in the
outer disc regions, interpreted in the framework of our simple spectral
model as the signature of optically thin emission, can have a different
interpretation, namely, that the outer parts of the disc show a vertical
structure with a temperature inversion possibly caused by irradiation
from the hot, inner disc regions.

The maps show that the bright spot, expected to appear where the
accretion stream hits the disc, is strongest in the hottest states and
curiously in the intermediate state (bin 4). The bin 4 light curve in
Fig.~\ref{fig_lcvs} shows a dip during egress at phase 0.08 and a kink
at phase $-0.04$ that must be responsible for the spot in the
reconstructed map. It is unlikely that this spot is due to an
unfortunate contamination of flickering, because 10 light curves
contributed to this bin and any such arbitrary effect should have been
averaged out.

The radial distribution of the effective temperature $\Teff(R)$
follows the steady state distribution only in the outer parts of the
disc. In the inner, optically thick, part of the
disc $\Teff(R)$ flattens out, irrespective of the sub-state. This
flattening in the innermost part with radii $R<0.1\Rl$ can only partly
be attributed to the MEM smearing (see VHH). BSH also found a
flattening in their maps of UU~Aqr and Rutten, van Paradijs \&
Tinbergen~(1992) in SW~Sex and V1315~Aql. It appears to be a feature
of SW~Sex stars. However, another interpretation is that the Balmer
Jump in the inner part could be in absorption. Since our simple model
cannot handle this, the effective temperatures in the inner disc
and subsequently also the mass accretion rates might be
underestimated. 

The mass accretion rate in the outer regions of the disc varies
between $5\times10^{16}$\gs\ in the highest sub-state (1 and 2) via
$3.5\times10^{16}$\gs\ in the intermediate sub-states (3 to 5) and
$2.5\times10^{16}$\gs\ in the lowest (6). However, in the bright spot
it can reach values of up to $3\times10^{17}$\gs.  Since the effective
temperature distribution is rather flat in the inner disc, it drops to
about $4\times10^{15}$\gs\ in the innermost part of the disc in all
sub-states.

Fig.~\ref{fig_ti1} and \ref{fig_ti2} also show the critical effective
temperature $\TeffA(R)$ derived from the critical mass accretion rate
$\Md_{\rm A}(R)$ as given by Ludwig et al.\ (1994). In nova-likes the
effective disc temperatures should lie above this critical value as to
prevent any dwarf nova-type outbursts. The plots show that the inner
optically thick part of the disc is at all times well above this
critical limit. However, the outer parts fall partly below, especially
in regions away from the bright spot. This might be the reason for the
mini-outbursts observed by Honeycutt et al.\ (1998).

The size of the optically thick region (see also
Section~\ref{sec_radii}), the presence of the bright spot and the mass
accretion rate suggest that in sub-states 1 and 4 the disc is most
active.  It is possible that the activity starts at the bright spot
and only slowly works itself through to the inner parts of the disc
(\#8518 to \#8524 and \#8030 to \#8042) or that the disc emission
drops suddenly (\#8512 to \#8518 and \#9025 to \#9031) without
immediately affecting the presence of the bright spot.

We performed PPEM on all individual light curves for comparison and
statistical uses. Images of the intensity ratio maps and radial
effective temperature profiles can be viewed at
http://pin\-guin.ast.uct.ac.za/$\sim$son\-ja/uuaqr\-maps.html.  A
postscript file of the maps can also be downloaded.

In the following Sections we focus on the white dwarf temperature, the
uneclipsed component and the disc radius as reconstructed for various
sub-states.

\subsection{Various aspects of the reconstructions}
\subsubsection{The white dwarf}
\label{sec_wd}

%%%%%%%%%%%%%%%%%%%%%%%%%%%%%%%%%%  TABLE  %%%%%%%%%%%%%%%%%%%%%%%%%%%%%%%%%%%%
\begin{table}[ht]
   \centering
\caption{The white dwarf temperature in the various sub-states of the
accretion disc $T_{wd,sub}$ and the average of the reconstructed white
dwarf temperatures derived for the individual light curves
$<T_{wd}>$. The error for $T_{wd,sub}$ must be estimated to about
1000~K, the standart deviation for $<T_{wd}>$ is given in the last
column. The second column lists the no. of eclipses contributing to
each bin.
\label{tab_wd}}
\begin{center}
\begin{tabular}{ccccc}
Bin & No. of eclipses & $T_{wd,sub}$ & $<T_{wd}>$ & $\sigma(<T_{wd}>)$ \\ \hline
1 & 3 & 23000 & 24200 & 1300 \\ 
2 & 6 & 29200 & 24200 & 3500 \\ 
3 & 5 & 26600 & 27533 & 2900 \\ 
4 & 10 & 24300 & 23900 & 5700 \\ \hline 
5 & 4 & 28100 & 30300 & 8800 \\ 
6 & 2 & 22000 & 30600 & 1900 \\ 
\hline
\end{tabular}
\end{center}
\end{table}
%%%%%%%%%%%%%%%%%%%%%%%%%%%%%%%%%%%%%%%%%%%%%%%%%%%%%%%%%%%%%%%%%%%%%%%%%%%%%%%

The white dwarf temperature $T_{wd}$ as reconstructed for the light
curves in the various sub-states is given in Tab.~\ref{tab_wd}. We
also list an average of the white dwarf temperatures reconstructed
after applying the PPEM method to the individual light curves. In
principle these temperatures should be roughly identical, however,
flickering and noise in the individual light curves affect the
resulting individual $T_{wd}$'s. This is most likely the case for the
two light curves in bin 6.  Furthermore, we could not always fit the
light curves to the same $\chi^2$ without introducing severe
artifacts. This might also influence the exact value of the
reconstructed $T_{wd}$'s, as probably happened in bin 2. We therefore
rely more on the values of $T_{wd}$ for the various substates rather
than the individual $T_{wd}$'s.  The error of the $T_{wd,sub}$ is
about 1000~K. This is determined by the change in $\chi^2$ due to an
artificial change of the white dwarf temperature in the otherwise
original reconstructed maps.

The $T_{wd}$'s derived for the four highest sub-states show a minimum
for the lowest and highest ones.  If one would assume the four
sub-states to represent a time sequence, this could be understood as a
reaction of the white dwarf temperature due to delayed accretion onto
the white dwarf: In the hottest state the white dwarf temperature is
practically identical to the low state temperature (sub-state 6). The
accretion disc, however, experiences enhanced accretion into the
bright spot, as evident from the prominent bright spot and reacts
rather quickly. The white dwarf can only slowly react to this change
in the accretion disc. Only in the second sub-state can the enhanced
accretion onto the white dwarf be seen, the white dwarf temperature is
at its maximum. For the next two bins the white dwarf cools down,
however, the enhanced accretion during bin 4 leads to another white
dwarf temperature maximum in bin 5. Bin 6 represent low states in
which the disc and the white dwarf are at their lowest activity.

As beautiful as this scenario looks, there are, however, a few
problems. First of all, the timescale for a change in the
temperature of the white dwarf is too short. The white dwarf cannot
cool down by a few 1000~K within a few orbits (less than a day) as
would be required for both jumps from sub-state 1 to 4 for eclipses
\#8512 to \#8518 and for \#9025 to \#9031 within 6 orbits.

Secondly, checking the light curves, the system does not always seem
to cycle through all sub-states (e.g.\ cycling from 4 to 3 and back:
\#8518 $\rightarrow$ \#8524 $\rightarrow$ \#8525 $\rightarrow$
\#8530), or seems to spend some time in one intermediate sub-state
before it suddenly jumps into a higher one (e.g.\ it rests in substate
2 during eclipses \#8586 and \#8591 before it jumps within one orbit
to sub-state 1 (\#8592)).

Since the sequences are unfortunately not taken continuously or with a
constant time intervall it is difficult to determine how exactly the
system behaved. Explicitly, we cannot always determine whether the
system was on its way into a high state or low state. In this regard,
it would be of great interest to gather light curves for several
consecutive nights.

It seems more likely that the actual location for the variable source
is the boundary layer around the white dwarf, since it is unlikely
that the white dwarf can cool so quickly. The fastest cooling rate
quoted by Sion (1999) is about 400-500 K per day (for RX~And) instead of
a few thousand Kelvin per orbit (4$h$) as would be necessary for
UU~Aqr. Although PPEM has a better constraint to determine the spatial
distribution of physical parameters, we am still left with some
ambiguity. The maximum entropy also prefers a solution with one bright
image pixel (e.g.\ the white dwarf) to more structure distributed over
a range of pixels, e.g.\ the inner part of the disc. Furthermore, the
white dwarf pixel value is not smeared out with the surrounding disc
pixels (VHH) which which will enhance this effect. The idea of a
variable boundary layer also supports the consequences we find for the
uneclipsed component as described in the following section.

The white dwarf temperatures are in general compatible with Sion's
(1999) results who finds an average white dwarf $\Teff$ of 24,100~K
for non-magnetic systems.

\subsubsection{The uneclipsed component}
\label{uneclipsed}

%%%%%%%%%%%%%%%%%%%%%%%%%%%%%%%%%%  TABLE  %%%%%%%%%%%%%%%%%%%%%%%%%%%%%%%%%%%%
\begin{table}[ht]
   \centering
\caption{Uneclipsed component in the filters UBVRI for the
reconstruction of the 6 binned light curves. The last two rows give
the expected flux of a dwarf of the indicated type at the distance of 205~pc.
\label{tab_ue}}
\vspace{1ex}
\begin{tabular}{cccccc}
 Bin/Flux(mJy)  & U & B & V & R & I\\ \hline
1 & 1.44 & 2.21 & 0.79 & 0.83 & 1.01\\
2 & 1.69 & 2.69 & 1.04 & 0.87 & 1.15\\
3 & 2.56 & 3.39 & 1.91 & 1.84 & 2.34\\
4 & 1.56 & 2.19 & 0.66 & 0.73 & 0.78\\ \hline
5 & 1.36 & 1.80 & 0.26 & 0.24 & 0.11\\
6 & 2.21 & 2.45 & 1.04 & 1.04 & 1.43\\ \hline
M4V &  --  &  --  & 0.06 & 0.21 & 0.81\\
M6V &  --  &  --  & 0.002 & 0.01 & 0.09\\
\hline
\end{tabular}
\end{table}
%%%%%%%%%%%%%%%%%%%%%%%%%%%%%%%%%%%%%%%%%%%%%%%%%%%%%%%%%%%%%%%%%%%%%%%%%%%%%%%

%%%%%%%%%%%%%%%%%%%%%%%%%%%%%%%%%%  TABLE  %%%%%%%%%%%%%%%%%%%%%%%%%%%%%%%%%%%%
%\begin{table*}[hbt]
%   \centering
%\caption{Averaged values of the uneclipsed component in the filters
%UBVRI derived from the individual light curves.
%\label{tab_ue2}}
%\vspace{1ex}
%\begin{tabular}{ccccccccccc}
% Bin/Flux(mJy)  & U & $\sigma_U$ & B & $\sigma_B$ & V & $\sigma_V$ & R & $\sigma_R$ & I & $\sigma_I$\\ \hline
%1 & 1.20 & 0.22 & 2.42 & 0.31 & 0.91 & 0.12 & 1.10 & 0.34 & 1.52 & 0.32\\
%2 & 1.59 & 0.42 & 2.71 & 0.42 & 1.20 & 0.38 & 1.21 & 0.30 & 1.74 & 0.32\\
%3 & 2.34 & 0.50 & 3.64 & 0.51 & 1.94 & 0.41 & 1.79 & 0.33 & 2.15 & 0.48\\
%4 & 1.52 & 0.30 & 2.40 & 0.23 & 0.74 & 0.20 & 0.98 & 0.23 & 0.93 & 0.24\\
%5 & 1.01 & 0.20 & 1.82 & 0.25 & 0.27 & 0.16 & 0.39 & 0.11 & 0.34 & 0.16\\
%6 & 2.03 & 0.25 & 2.71 & 0.25 & 1.04 & 0.03 & 0.72 & 0.16 & 1.12 & 0.14\\
%\hline
%\end{tabular}
%\end{table*}
%%%%%%%%%%%%%%%%%%%%%%%%%%%%%%%%%%%%%%%%%%%%%%%%%%%%%%%%%%%%%%%%%%%%%%%%%%%%%%%

%%%%%%%%%%%%%%%%%%%%%%%%%%%%%%%%%% FIGURE %%%%%%%%%%%%%%%%%%%%%%%%%%%%%%%%%%
\begin{figure*}
   \centering
\includegraphics[width=58mm]{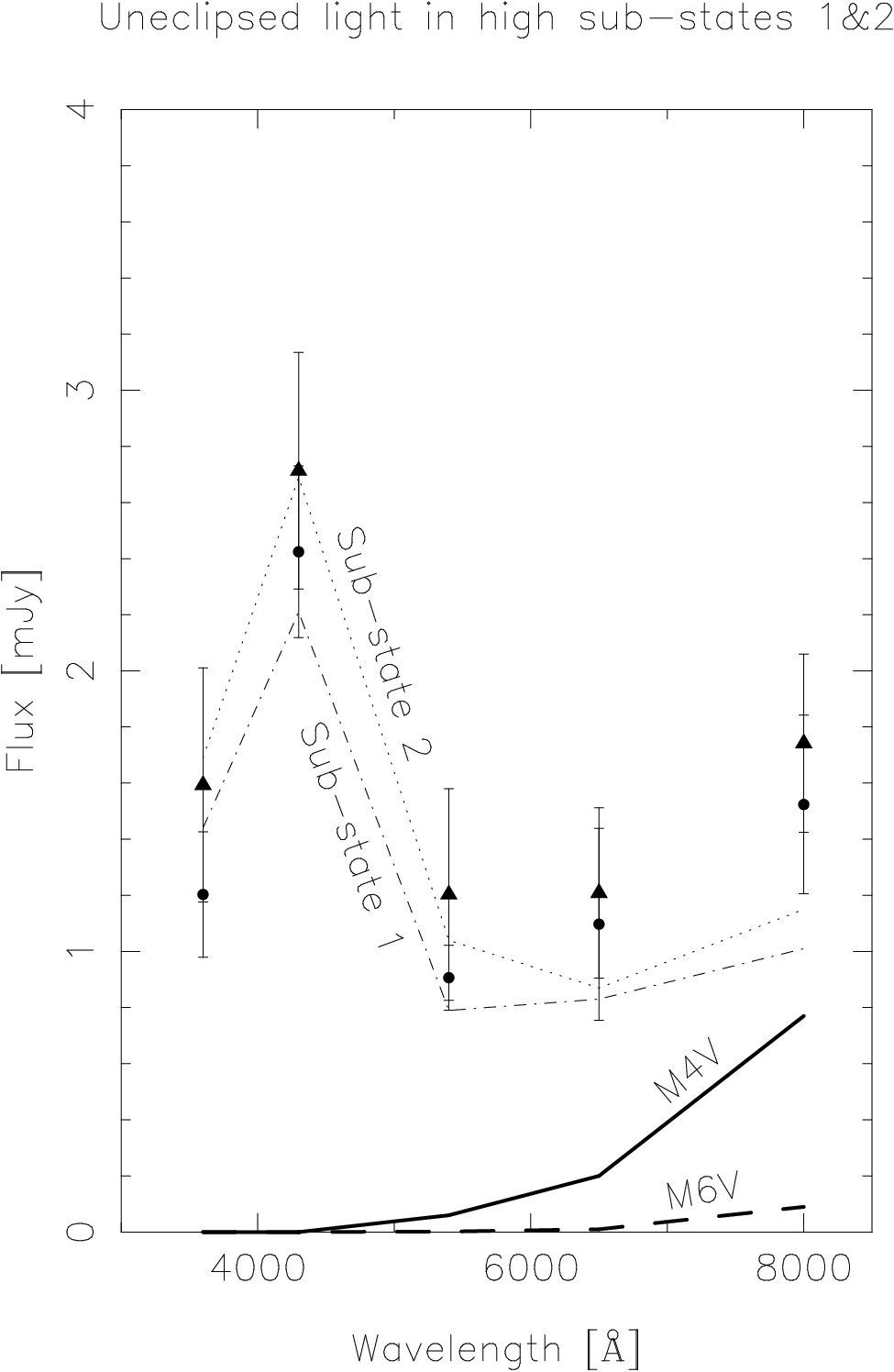}
\includegraphics[width=58mm]{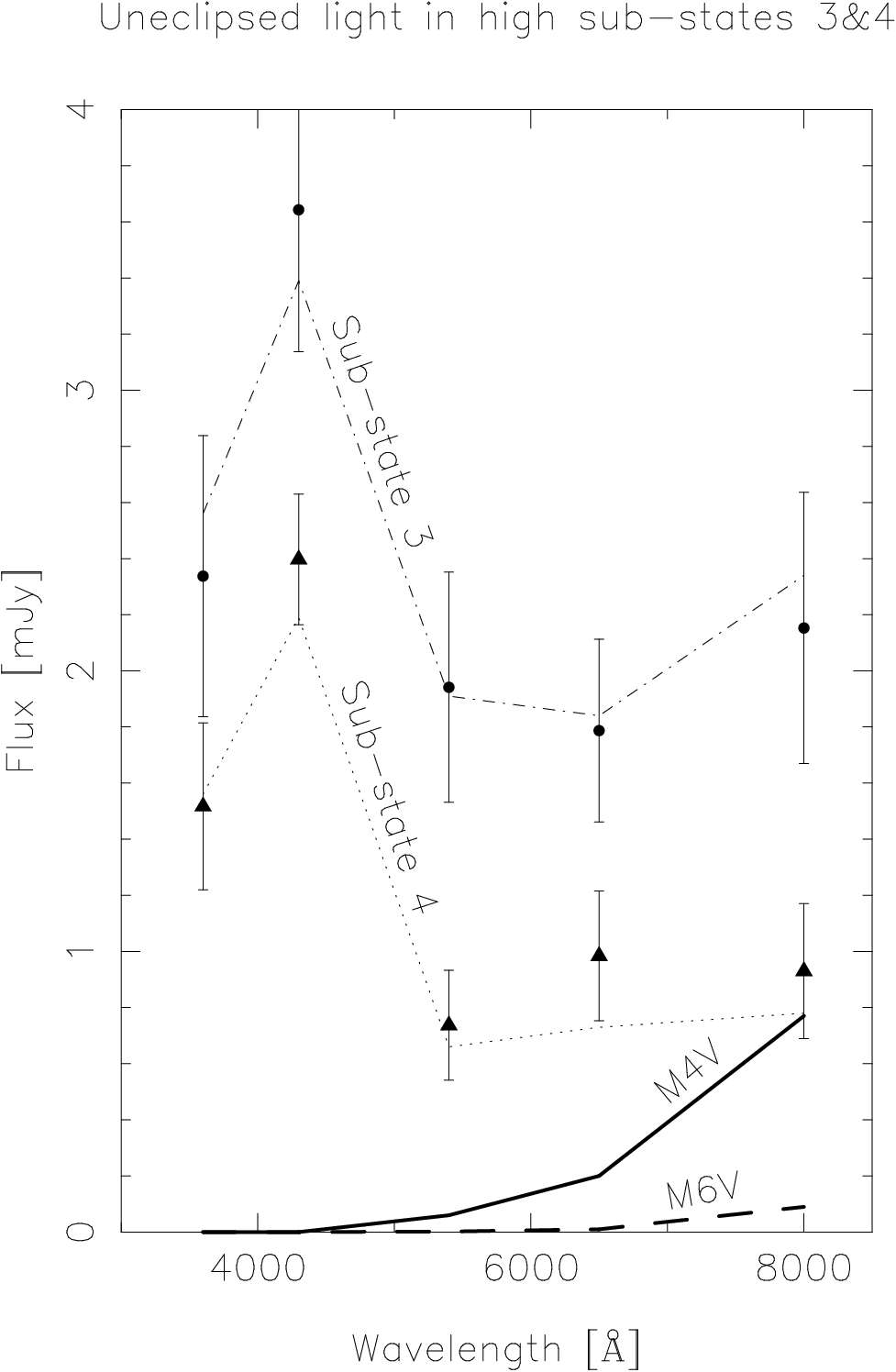}
\includegraphics[width=58mm]{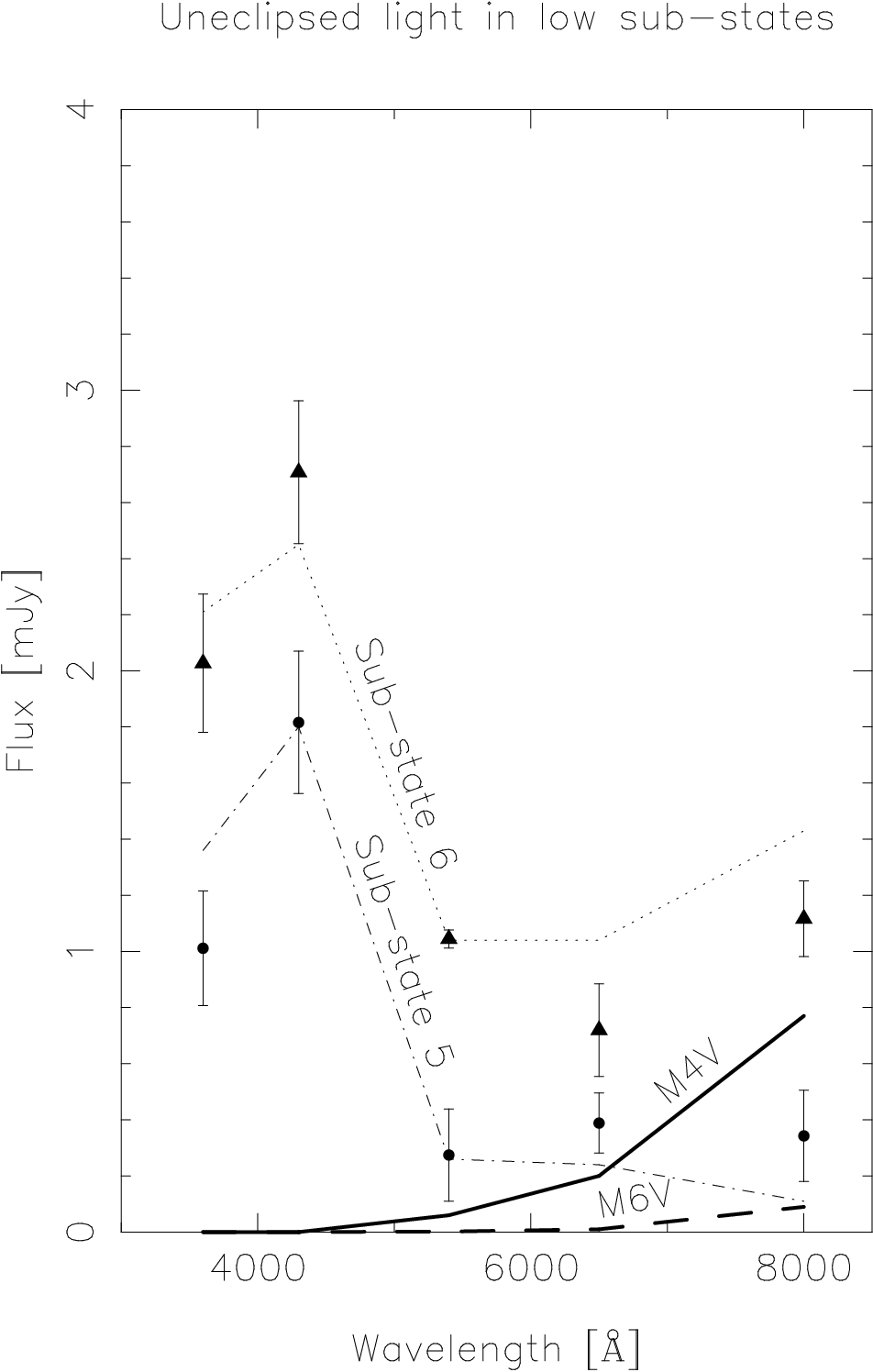}

\caption{The uneclipsed component in the various sub-states as
indicated in the graphs (thin lines). The points with error bars (full
circles: sub-states 1,3,5; full triangles: sub-states 2,4,6,
respectively) give the average of the uneclipsed spectra from the
individual reconstructions and are largely compatible with the
sub-state uneclipsed spectra. Furthermore, the error bars give an
indication of the reliability of the sub-state spectra. For comparison
the flux of M4V (thick solid line) and M6V (thick dashed line) dwarf
stars are drawn at the bottom of each graph.}
\label{fig_unecl}
\end{figure*}
%%%%%%%%%%%%%%%%%%%%%%%%%%%%%%%%%% FIGURE %%%%%%%%%%%%%%%%%%%%%%%%%%%%%%%%%%

The uneclipsed component for the various filters and bins is given in
Tab.~\ref{tab_ue} and Fig.~\ref{fig_unecl}.  The uneclipsed flux
consists of the flux from the secondary and other sources in the
system that are never eclipsed, like a disc wind or a chromosphere as
suggested by Baptista et al.\ (2000).

The secondary has a mass of $0.20\pm0.07\Msol$ (BSC) which fits to a
M(4$\pm$1)V dwarf star according to Kirkpatrick \& McCarthy
(1994). Table~\ref{tab_ue} gives the flux of an M4V dwarf star at a
distance of 205~pc. If the secondary is a M4V dwarf, then the
reconstructed I-band flux in sub-state 5 is too small. However, the
test in Appendix~\ref{app_test} shows that the uneclipsed flux in all
filters might be underestimated. Furthermore, the CVs analysed by
Beuermann et al. (1998) with an orbital period similar to that of
UU~Aqr have a spectral type of M$4\pm1$.

Another constraint for the spectral type of the secondary comes from
the K light curve (Huber 2001, private communication). The
out-of-eclipse light curve ($m_K \sim 12$~mag) shows orbital variation
that can be attributed to the ellipsoidal change in surface area of
the secondary and an eclipse of the secondary by the outer edges of
the disc. The full amplitude is 0.14~mag or 13\% of the minimum
flux. An M4V dwarf has a K magnitude of 7.80 (Kirkpatrick \& McCarthy
1994) and would therefore contribute 13\% to the minimum K flux. If
the secondary has a spectral type of M4V this means that it were fully
eclipsed in order to cause the observed orbital variation. This cannot
be the case. Only if it is brighter than a typical M4V dwarf can
the ellipsoidal shape cause such strong orbital variation. The
secondary should therefore have an earlier spectral type than M4V.

Our estimated limit for the secondary of 13\% of the K flux is
compatible with Dhillon et al.'s (2000) upper limit of 55\% derived
from the lack of absorption features in the IR spectra.

The remaining part of the uneclipsed component must arise in other
parts of the system.  As we show in Appendix~\ref{app_test} the excess
in the uneclipsed B flux is an artifact due to the 
spectral model we used with PPEM. The cause is most likely that the
size and optical depth of the bright spot is influenced by maximum
entropy smearing.  Disregarding the strong B-band flux the uneclipsed
spectrum could be compatible with optically thin emission or more
likely of an extended source with varying temperature and
density. This could be from a disc chromosphere or a disc wind or a
combination of both as suggested by Baptista et al.~(2000).

A look at the changes of the uneclipsed contribution through the
various sub-states show that the uneclipsed component seems to react
even slower than the white dwarf temperature or boundary layer (if the
substates do represent a time sequence). Only in sub-state 3 there is
a significant increase in the uneclipsed component.

With the assumption that although the absolut flux units are
errorneous, the {\it shape} of the uneclipsed spectrum in the UVRI
passbands are represented correctly (see Appendix~\ref{app_test}), we
can make a cautious statement about the variation of the uneclipsed
spectrum with substate.  The difference in flux in the various high
sub-states appears to be constant with respect to wavelength except
for a slight increase in the I-band.  This is also true for the two
low states and means that the varying part of the uneclipsed component
either consists of a nearly wavelength independent plus a cool source
or a moderately hot plus a cool source. The cool source could be the
outer regions of a disc wind driven vertically out of the disc, as
suggested already by BSC and calculated by Pereyra \& Kallman
(2000). Any fluctuations in the disc wind will be most noticeable in
the outer, cooler regions of the disc wind.

IR spectra taken by Dhillon et al.\ (2000) show that the emission
lines profiles of B$\gamma$ (Brackett series of H~I), B$\delta$ and
He~I $\lambda 2.0587\mu$m are single-peaked. This is understandable if
the emission lines originating in the cool disc wind dominate over the
line emission from the disc. This disc wind is possibly triggered by
the changes in the mass accretion rate.

\subsubsection{The disc radius}
\label{sec_radii}

%%%%%%%%%%%%%%%%%%%%%%%%%%%%%%%%%% FIGURE %%%%%%%%%%%%%%%%%%%%%%%%%%%%%%%%%%
\begin{figure}
   \centering
\includegraphics[width=80mm]{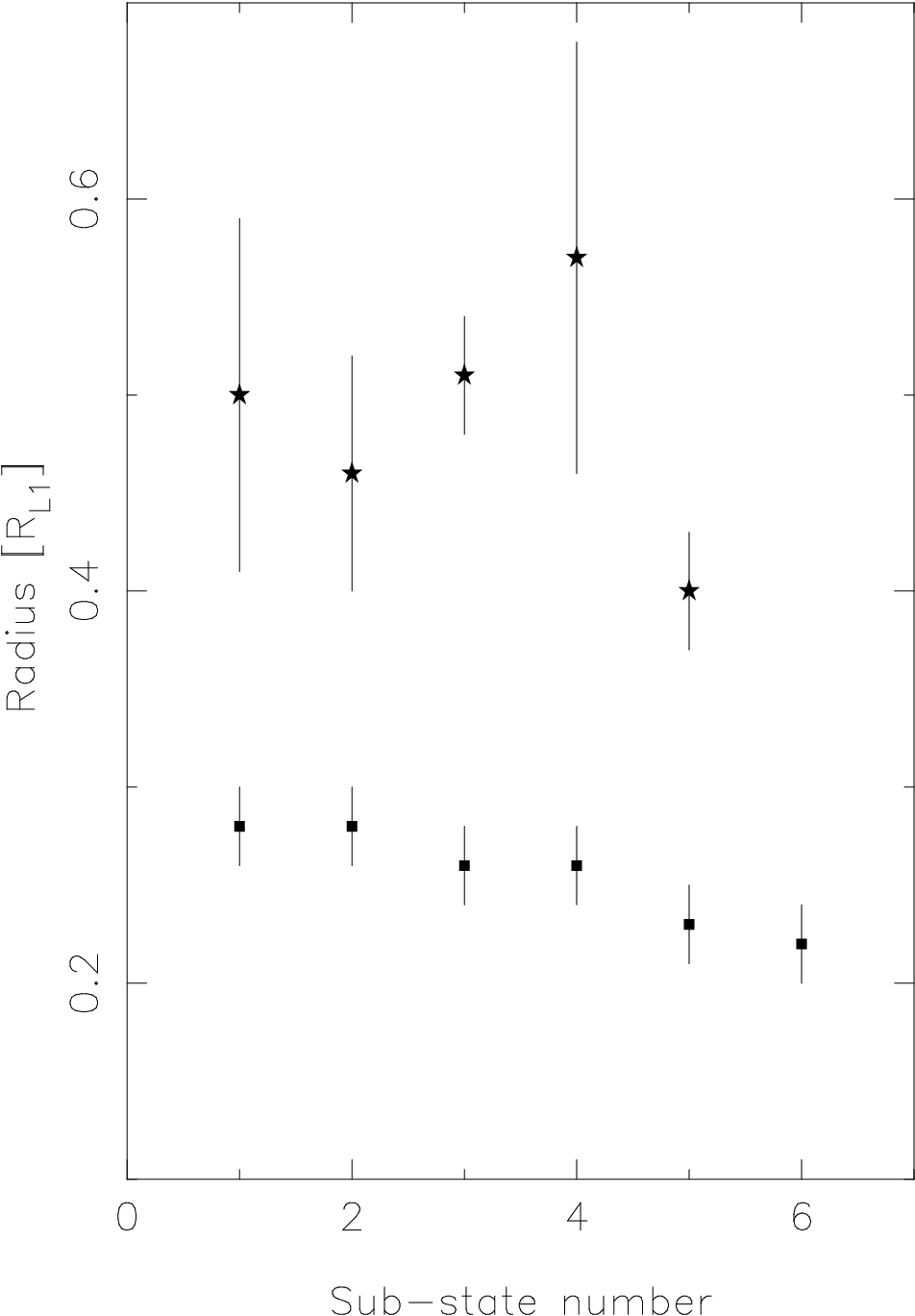}
\caption{The radius of the optically thick part of the disc (squares
with error bars) and the radius of the bright spot (asterixes with
error bars). Th bright spot radii must be increased by about 0.07$Rl$,
judging from tests with artificial data. $1~\Rl =
5.2\times10^{10}$~cm.}
\label{fig_rad}
\end{figure}
%%%%%%%%%%%%%%%%%%%%%%%%%%%%%%%%%% FIGURE %%%%%%%%%%%%%%%%%%%%%%%%%%%%%%%%%%

%%%%%%%%%%%%%%%%%%%%%%%%%%%%%%%%%%  TABLE  %%%%%%%%%%%%%%%%%%%%%%%%%%%%%%%%%%%%
\begin{table}[ht]
   \centering
\caption{Radii of features in the accretion disc: $R_{cen}$ is the
central optically thick region (errors are 0.02$\Rl$), $R_{spot}$ the
radius of the bright spot (tests show these values must be
increased by about 0.07$\Rl$) and $R_{I}$ an estimate of the disc radius using
the intensity distribution in the I-band filter (errors are
0.05$\Rl$). The latter is an upper limit for the disc size.
\label{tab_rad}}
\vspace{1ex}
\begin{tabular}{cccccc}
 Bin  & $R_{cen}$ & $R_{spot}$ & $\sigma(R_{spot})$ & $R_{I}$ \\ \hline
1 & 0.28 & 0.50 & 0.09 & 0.71 \\
2 & 0.28 & 0.46 & 0.06 & 0.61 \\
3 & 0.26 & 0.51 & 0.03 & 0.66 \\
4 & 0.26 & 0.57 & 0.11 & 0.71 \\ \hline
5 & 0.23 & 0.4  & 0.03 & 0.57 \\
6 & 0.22 &  --  &  --  & 0.63 \\
\hline
\end{tabular}
\end{table}
%%%%%%%%%%%%%%%%%%%%%%%%%%%%%%%%%%%%%%%%%%%%%%%%%%%%%%%%%%%%%%%%%%%%%%%%%%%%%%%

%\subsubsection{The measurements}
From the maps in Fig.~\ref{fig_ti1} and \ref{fig_ti2} we measure the
radius of the optically thick part of the disc and the bright spot,
where applicable, i.e.\ for all but the last bin. The values are
listed in Table~\ref{tab_rad}. 

A test with artificial data shows (Appendix~\ref{app_test}) that the
spot location is reconstructed at a slightly larger azimuth (judging
from the brightness maximum in the spot) and a slightly smaller
radius, by about 0.07$\Rl$ (this value decreases with decreasing
$\chi^2$ and increasing signal-to-noise factor). The true spot
locations will thus be at a slightly larger radius than given in
Table~\ref{tab_rad}.

The spot radius increases from sub-state 1 to 4. However, since the
error bars are rather large, a constant spot radius of about
$R_{spot}(high) = 0.51\Rl \pm 0.04\Rl (= (2.6\pm0.2)\times10^{10}$~cm)
is also compatible with the data.  Only in the sub-state 5 does the
spot radius decrease slightly to $R_{spot}(low) = 0.40\Rl (=
2.1\times10^{10}$~cm).

The spot radius should be a good measure of the disc radius. However,
due to the lack of any traces of the bright 
spot in the lowest sub-state we unfortunately cannot determine a
comparable measure for the disc size. Analysing the intensity maps we
can only estimate an upper limit of the disc radius, as given in
Table~\ref{tab_rad}.

Fig.~\ref{fig_rad} also gives the radius of the transition zone
between the inner optically thick accretion disc and the optically
thin outer parts (or the radius at which the temperature inversion in
the outer disc becomes dominant). It decreases from $0.28\Rl$ in the highest
sub-states to $0.22\Rl$ in the lowest. Baptista et al. (2000) find a
transition zone at between $0.2\Rl$ and $0.3\Rl$ in which the emission
lines show P Cygni profiles for which either the absorption (H$\delta$,
He~I~$\lambda$5876) or emission (H$\gamma$, H$\beta$, H$\alpha$)
component is stronger. Their result is therefore compatible with ours.

\subsection{Discussion}

\subsubsection{The accretion disc}

Our disc radii ($R_{spot}(high) = 0.51\Rl (+ 0.11\Rl - 0.04\Rl)$ and
$R_{spot}(low) = 0.40\Rl (+ 0.1 \Rl - 0.03\Rl)$) are somewhat smaller,
but compatible with previously derived values. Harrop-Allin \&
Warner (1996) find disc radii of UU~Aqr of $0.61<R_d/\Rl<0.88$ in the
high state and $0.48<R_d/\Rl<0.72$ in the low state while BSC find
$\sim 0.72\Rl$ and $\sim0.38\Rl$, respectively.  The eclipse map in
BSH's Fig.~9b, however, shows the spot in the high state at
$0.56\pm0.12\Rl$, compatible with our estimate.  Harrop-Allin \&
Warner as well as BSC determined the disc size by the bright spot
contact phases and appear to give slightly too large results, while
Eclipse Mapping seems to derive too small ones. However, the radii
derived from eclipse mapping techniques can be improved with better
signal-to-noise ratio data and better fits (using more appropriate
spectral models) as tests show.  Furthermore, Eclipse
mapping methods use the information of the whole eclipse profile and
are therefore minimally interfered by flickering. On the other hand,
accretion stream overflow could in principle cause the bright spot to
appear closer to the white dwarf than the actual disc edge. For
example, Horne's (1999) model for SW~Sex stars includes mass overflow.

Since the disc radius does not seem to change with the variations
during high state, the inner optically thick disc must be directly
responsible for the changes in the out-of-eclipse flux. The variation
in the size of the optically thick part of the disc is presumably
directly correlated to a redistribution of the mass.

UU~Aqr's disc is so close to the critical mass accretion rate that it
is not surprising that the system appears so variable as already
pointed out by BSH. It is not only switching between low to high
states, but also varying in the high state. Furthermore, it shows
stunted outbursts during low state as observed by Honeycutt et al.\
(1998).  Our analysis suggests that these stunted outbursts might be
caused in the outer disc in which the temperatures fall below the
critical value. If this is correct they are due to a disc instability
like normal outbursts in dwarf novae, however, shorter and with lower
amplitude, because the temperatures in the outer disc easily reach the
critical values and the temperatures in the inner disc are already
above the critical limit. The cause of the variation between high and
low state, however, is more likely a varying mass accretion rate
as suggested by BSH. It would be very valuable to
know if the normal outbursts as indicated by Volkov et al.'s (1986)
observations start during high or low state and how frequently they
occur.

\subsubsection{The disc wind}

Baptista et al.~(2000) give a detailed discussion on the presence of a
disc wind and chromosphere in UU~Aqr (we do not want to repeat their
argumentation). Our observational results are compatible with
this most likely scenario. Our results also favour a geometrically
extended (maybe spherical or equatorial) source that could be an
outflowing gas and/or chromosphere covering areas of different
temperatures rather than a collimated jet.

In Doppler tomographs a disc wind with a velocity of $v_w$ would
appear as a ring with a diameter of the projected velocity ($v_w\cos
i$). (In high resolution Doppler maps one could possibly detect a
small hole in this ring at $(v_x = v_w\cos i, v_y = 0)$ due to the
exclusion of the eclipse phases.)  Since the inclination angle $i$ is
large, this ring would be small and occupy the central regions of the
Doppler maps. The Doppler maps of Kaitchuck et al.~(1998) show
emission from the central regions apart from the accretion disc and
bright spot contributions.  While a clear ring-like structure cannot
be seen, the half-orbit maps support such emission feature. Improved
signal-to-noise spectra are highly desireable in order to decide about
this type of feature in the Doppler maps.

\begin{acknowledgements}
We thank Brian Warner, Stephen Potter, David Buckley and Encarni
Romero Colmenero (the Binary Accretion Natter Group) for fruitful
discussions. Thanks go also to the unknown referee for helpful
comments leading to a clearer presentation of the paper.

\end{acknowledgements}

\begin{appendix}
%%%%%%%%%%%%%%%%%%%%%%%%%%%%%%%%%% FIGURE %%%%%%%%%%%%%%%%%%%%%%%%%%%%%%%%%%
\begin{figure*}
   \centering
   \includegraphics[width=67mm]{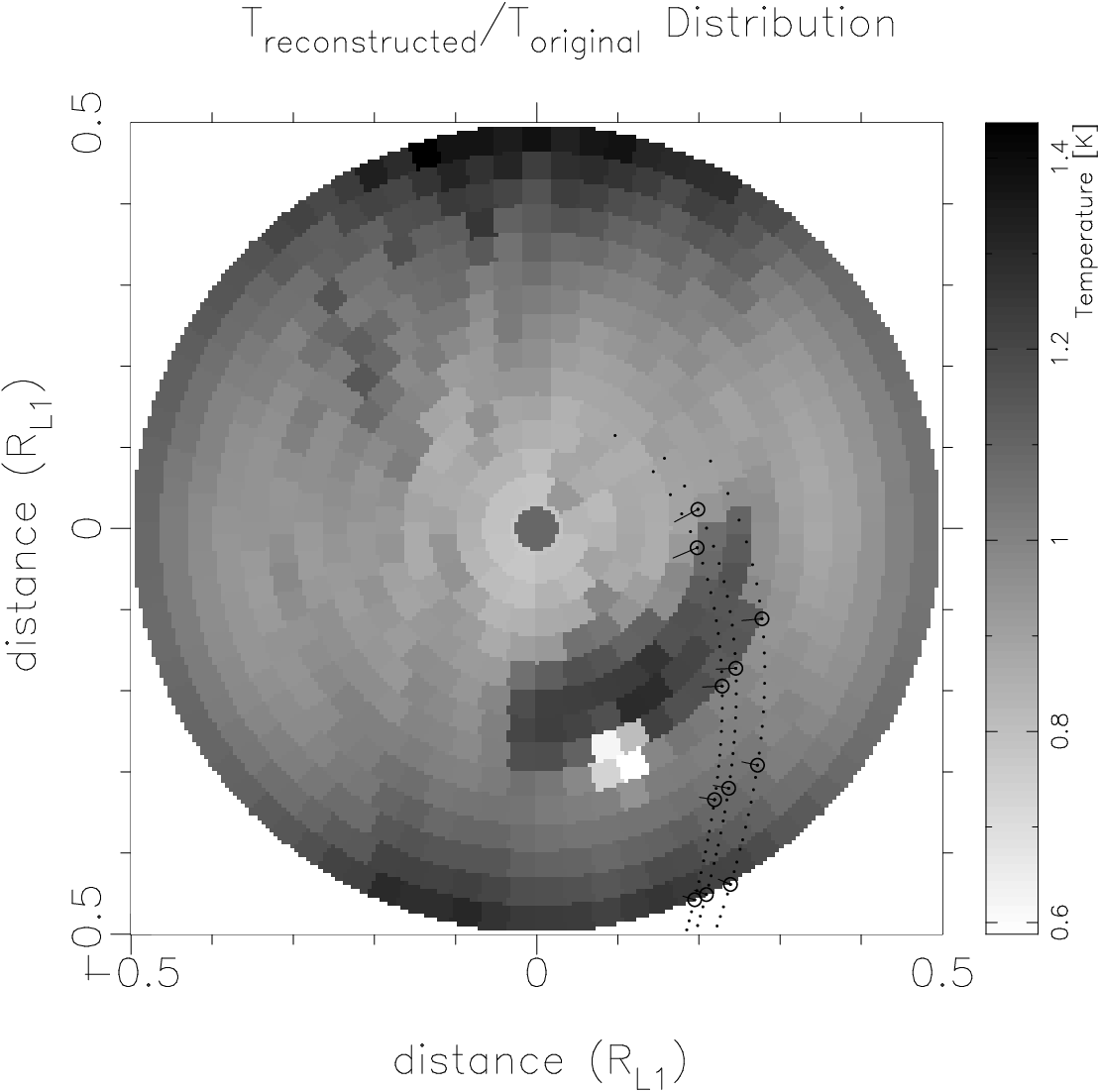}
\hspace{1cm}
   \includegraphics[width=67mm]{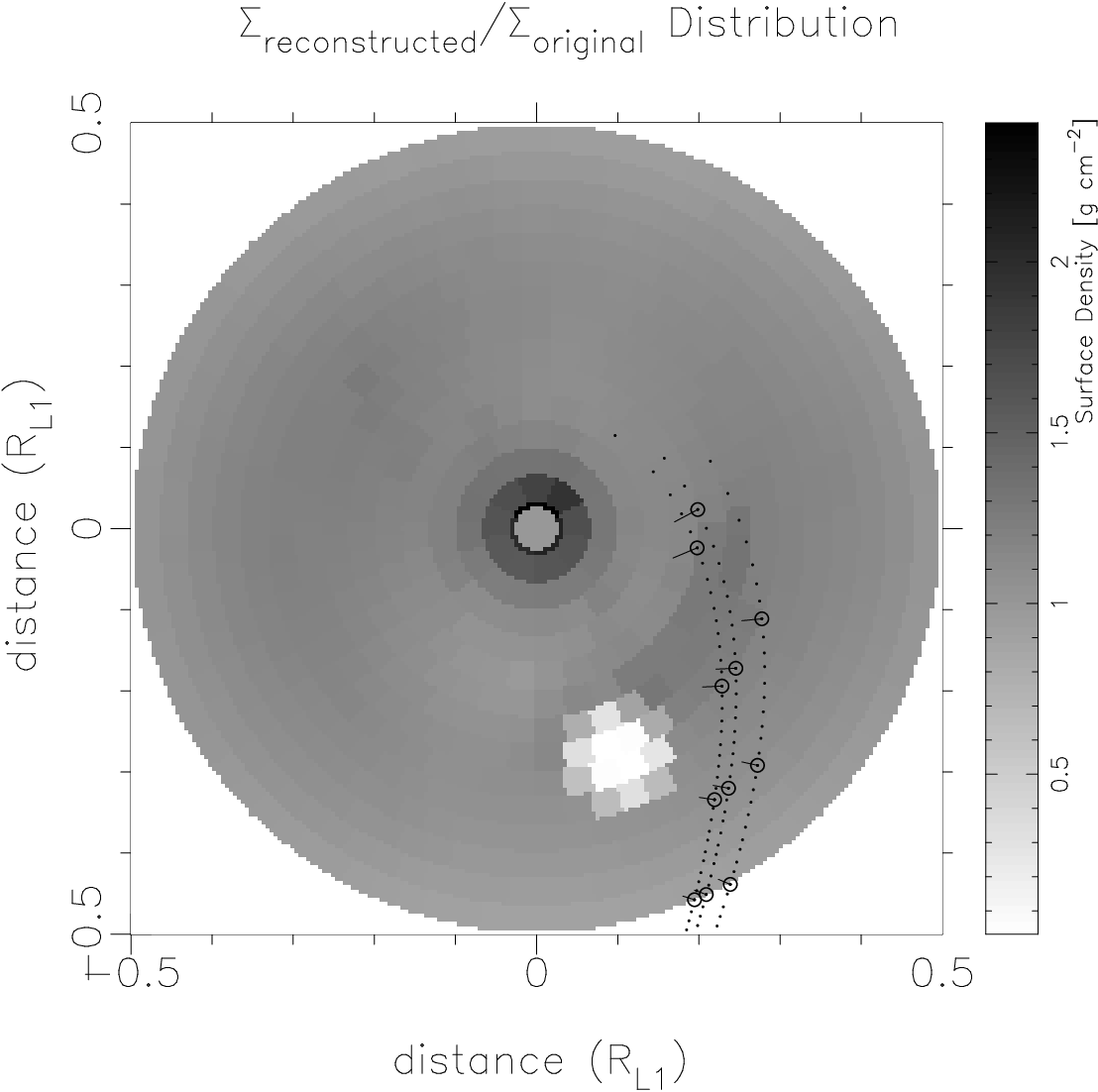}
\caption{Ratio of reconstructed to original parameter maps for the
test case. The bright area indicates the location where the original
spot was placed, the dark, slightly overlapping region where spot was
reconstructed.}
\label{testv}
\end{figure*}
%%%%%%%%%%%%%%%%%%%%%%%%%%%%%%%%%% FIGURE %%%%%%%%%%%%%%%%%%%%%%%%%%%%%%%%%%

\section{Tests with artificial discs}
\label{app_test}

%%%%%%%%%%%%%%%%%%%%%%%%%%%%%%%%%%  TABLE  %%%%%%%%%%%%%%%%%%%%%%%%%%%%%%%%%%%%
\begin{table}[ht]
   \centering
\caption{Flux of the original and reconstructed uneclipsed component in the
filters UBVRI for the test case (in mJy).
\label{tabtest_ue}}
\vspace{1ex}
\begin{tabular}{cccccc}
         & U & B & V & R\\ \hline
original & 0.09 & 0.069 & 0.69 & 0.16 \\
reconstructed & 0.04 & 0.03 & 0.00 & 0.07 & \\
\hline
\end{tabular}
\end{table}
%%%%%%%%%%%%%%%%%%%%%%%%%%%%%%%%%%%%%%%%%%%%%%%%%%%%%%%%%%%%%%%%%%%%%%%%%%%%%%% 

%%%%%%%%%%%%%%%%%%%%%%%%%%%%%%%%%% FIGURE %%%%%%%%%%%%%%%%%%%%%%%%%%%%%%%%%%
\begin{figure}
   \centering
   \includegraphics[width=70mm]{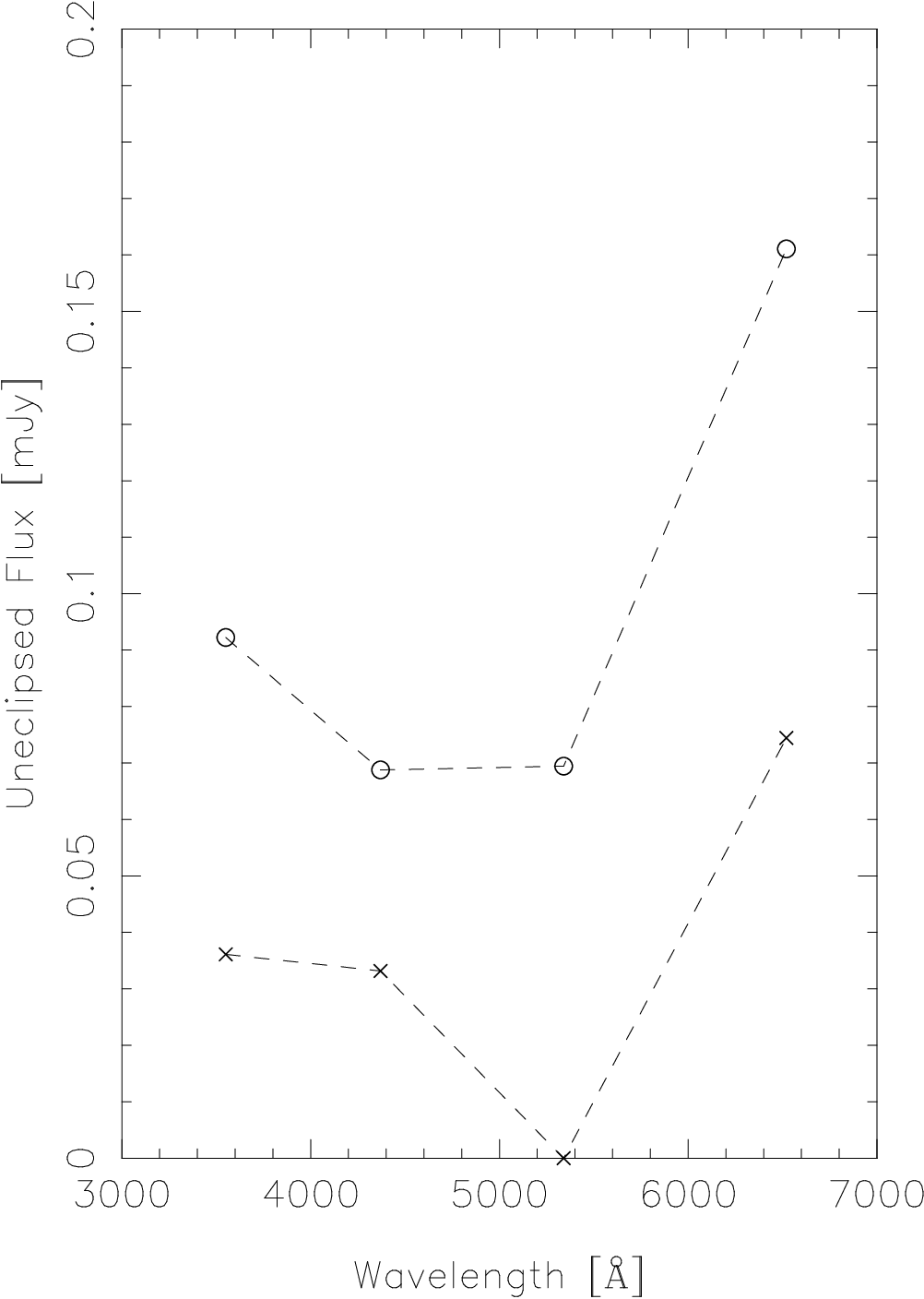}
\caption{Uneclipsed component for the original (circles) and
reconstructed (crosses) accretion disc in the test case.}
\label{testue}
\end{figure}
%%%%%%%%%%%%%%%%%%%%%%%%%%%%%%%%%% FIGURE %%%%%%%%%%%%%%%%%%%%%%%%%%%%%%%%%%

In order to explain the excess in the B-band flux we performed a test of the
PPEM method with artificial data. We constructed an axisymmetric disc
with a gaussian bright spot at a radius of $0.3\Rl$, a sigma of
0.02$\Rl$ and an azimuth of 20$^\circ$, calculated the light curve in
UBVR with 112 data points between $-0.1$ and 0.1. To this light curve we
added a signal-to-noise of 50 and a constant factor of 0.01~mJy to
the error bars in order to keep a minimum error for small fluxes,
e.g.\ during eclipse. The uneclipsed component was set artificially to
an optically thin spectrum.

This light curve was the input for the PPEM programme and could be
fitted to a $\chi^2$ of 1.1. Fig.~\ref{testv} shows the ratios of the
reconstructed map to the originals, highlighting the locations of the
spot. The reconstructed spot is located  roughly at the right
position, but at a slightly smaller radius of $0.23\Rl$ and slightly
larger azimuth of 30$^\circ$ than the original one. Furthermore, the
spot is smeared out due to the maximum entropy (MEM) constraint.
Such a spreading of the spot could also be realistic, so
in reconstructions of real data one cannot give a precise size for the
azimuthal extension of the spot.

The uneclipsed component was reconstructed as shown in
Table~\ref{tabtest_ue} and Fig.~\ref{testue}. Most striking is the
fact that the reconstructed flux is lower than the original one. This
is due to the MEM smearing leading to larger maps and/or smoother
gradients in the outer parts of the disc. This leads to more flux in
the outer, uneclipsed regions of the disc.

Except for the B-band flux, the general shape of the uneclipsed spectrum is
recovered relatively well. The reason for the excess in the
reconstructed B-band flux in due to the smearing of the spot, leading to
larger parts of the disc being (nearly) optically thick. Our model
spectrum therefore underestimates the B-band flux in the disc, leading to
an increase in the uneclipsed component in order to fit the
out-of-eclipse flux.

\end{appendix}
\end{document}